\documentclass[aps,prd,superscriptaddress,nofootinbib,twocolumn]{revtex4-2}%
\usepackage{graphicx}
\usepackage{amsmath}
\usepackage{amssymb}
\usepackage{bm}
\usepackage{epsfig}
\usepackage{listings}
\begin{document}

\title{
Color-octet nonrelativistic QCD matrix elements for heavy quarkonium decays 
in the refined Gribov-Zwanziger theory
}
\author{Hee~Sok~Chung}
\affiliation{Department of Physics, Korea University, Seoul 02841, Korea}
\email{neville@korea.ac.kr}



\date{\today}

\begin{abstract}
We determine color-octet nonrelativistic QCD matrix elements for quarkonium
decays from moments of the two-point correlation function of the QCD
field-strength tensor computed in the refined Gribov-Zwanziger theory. 
We find that a tree-level calculation in the refined Gribov-Zwanziger theory
can give a suitable description of the QCD field-strength correlation function
at both short and long distances, which leads to moments that are infrared
finite and can be properly renormalized. 
By using the color-octet matrix elements we obtain, we quantitatively improve
the nonrelativistic effective field theory description of quarkonium decay
rates, especially for the $\chi_{QJ}$ and $\eta_Q$ states, where $Q =c$ or
$b$.
\end{abstract}

\maketitle

\section{Introduction}
\label{sec:intro}

The two-point correlation function of the QCD field-strength tensor has been 
considered an important quantity in phenomenological studies of the strong
interaction~\cite{Shifman:1978bx, Voloshin:1978hc,
Leutwyler:1980tn, Dosch:1994wj, Dosch:1987sk, Dosch:1988ha, Nachtmann:1983uz,
Landshoff:1986yj, Kramer:1990tr, Dosch:1994ym, Gromes:1982su,
Campostrini:1986hy, Simonov:1995ui, Simonov:1988mj, DiGiacomo:2000irz}. 
In particular, it has been known that they can be used to compute
nonperturbative matrix elements that arise in decays of heavy quarkonium
states~\cite{Bodwin:1994jh, Brambilla:2001xy, Brambilla:2002nu}. 
The contributions to the decay rates of heavy quarkonia from probabilities 
for a heavy quark $Q$ and antiquark $\bar Q$ pair to be in a 
color-octet state are 
encoded in the nonrelativistic QCD (NRQCD) color-octet matrix elements, 
which can be expressed as products of quarkonium wave functions at the
origin and moments of field-strength correlators as have been shown in the
potential NRQCD (pNRQCD) effective field theory
formalism~\cite{Pineda:1997bj, Brambilla:1999xf, Brambilla:2001xy,
Brambilla:2002nu, Brambilla:2004jw}. 
The color-octet contributions can have significant effects on decay rates of
heavy quarkonia. Most notably, in inclusive decays of $P$-wave heavy quarkonia
the color-octet contribution appears at leading order in the nonrelativistic
expansion in powers of $v$, the typical velocity of the heavy quark inside the
quarkonium~\cite{Bodwin:1992ye}. 
Color-octet contributions also appear in two-photon decays of $P$-wave 
heavy quarkonia as corrections of order $v^2$~\cite{Brambilla:2017kgw}. 
Even in the case of $S$-wave quarkonium decays, 
inclusion of color-octet contributions are necessary for improving the theory 
description of decay rates and two-photon branching fractions of
$\eta_c$~\cite{Bodwin:2001pt, Brambilla:2018tyu}. 
It is also known that color-octet contributions are enhanced by an inverse
power of $\alpha_s$ in the case of $J/\psi$ and $\Upsilon$ 
decays~\cite{Petrelli:1997ge, Bodwin:2002cfe}. 
As the quarkonium wave functions at the origin can usually be determined from
potential models and electromagnetic decay rates of heavy quarkonia, 
and even be computed accurately from first principles~\cite{Chung:2020zqc,
Chung:2021efj}, 
quarkonium decays can provide useful probes of the moments of QCD
field-strength correlators. 

The moments of field-strength correlators are sensitive to both the
short-distance and long-distance behaviors of the correlators. 
The correlators as functions of the separation of the field strengths have been
studied in perturbative QCD~\cite{Eidemuller:1997bb}, 
in operator-product expansion~\cite{Braun:2020ymy, Braun:2021cqe}, 
and on the lattice~\cite{DiGiacomo:1992hhp, DElia:1997sdk, Bali:1997aj}.
The calculations in both perturbative QCD and operator-product expansion 
approaches only reproduce the
short-distance behaviors of the correlators, so they cannot be used to compute 
the moments; these calculations give results for the correlators that are
given by inverse powers of the separation of the two field strengths, 
whose moments are scaleless divergent and vanish in dimensional regularization. 
On the other hand, lattice QCD calculations have been shown to reproduce both
the nonperturbative long-distance behavior and the power-law behavior at short
distances. 
However, the short-distance behaviors of the lattice results are not well
understood in terms of perturbative QCD calculations, and this makes it 
difficult to renormalize the ultraviolet (UV) divergences in the moments. 
Renormalization of the moments is important because it is 
directly related to renormalization of the UV divergences in the
color-octet matrix elements. 
Hence, currently available results for the field-strength correlators do not
immediately lead to quantitative results for their moments. 
It would be desirable to have calculations for the correlators that are
compatible with perturbative QCD at short distances, and also
describe the nonperturbative long-distance behaviors at least approximately.

It has been known that non-Abelian gauge theories quantized in the standard
Fadeev-Popov method contain Gribov copies~\cite{Gribov:1977wm}; 
that is, a gauge-fixing condition
such as $\partial \cdot A = 0$, where $A$ is a gauge field, does not uniquely
fix the gauge and there are distinct gauge-field configurations that satisfy
the gauge-fixing condition that are related by large gauge transformations. 
In order to remove this ambiguity, the functional integral can be restricted to
a region free of Gribov copies called the fundamental modular region. 
A semiclassical calculation of the gluon propagator in the Landau gauge leads
to a result that is modified in the infrared region by a dimensionful quantity
called the Gribov parameter in a way that the poles of the propagator are
shifted to nonphysical locations, while the large-momentum behavior of the
propagator coincides with perturbative QCD. The Gribov parameter appears in the
modified propagator in the form of a complex-valued gluon mass, so that
infrared divergences are regulated while single-gluon states do not appear in
the physical spectrum. Later it was shown that by introducing auxiliary fields
the restriction to the fundamental modular region can be incorporated into 
a local and
renormalizable action, hereafter referred to as the Gribov-Zwanziger (GZ)
theory~\cite{Zwanziger:1989mf}, whose tree-level gluon propagator reproduces the semiclassical result
in Ref.~\cite{Gribov:1977wm}. We refer readers to Refs.~\cite{Gribov:1977wm, 
Henyey:1978qd, vanBaal:1991zw, Zwanziger:1989mf, Sobreiro:2004us} and the
review in Ref.~\cite{Vandersickel:2012tz} for details of the Gribov ambiguity
and the GZ theory. 
Because the tree-level propagator follows from a semiclassical calculation,
one would hope that it would at least approximately describe the
nonperturbative behavior of the gluon propagator. Unfortunately, lattice QCD
calculations have shown that this is not the case; while the tree-level gluon
propagator at zero momentum vanishes in the GZ theory, lattice data shows that
it is nonzero~\cite{Cucchieri:2004mf, Bowman:2007du}. 
This discrepancy suggests that more nonperturbative effects will need to be
included in order to describe the nonperturbative behavior of non-Abelian gauge
theories. 

The refined Gribov-Zwanziger (RGZ) theory has been obtained by adding 
effects of dimension-two condensates associated with the gauge field and the
auxiliary fields to the GZ action~\cite{Dudal:2007cw, Dudal:2008sp,
Dudal:2008rm, Dudal:2008xd, Dudal:2010tf, Dudal:2011gd}.
The dimension-two condensate of the gauge field has been considered an important
quantity in the study of non-Abelian gauge theories~\cite{Gubarev:2000eu,
Gubarev:2000nz, Verschelde:2001ia}; a gauge-invariant
definition of the dimension-two condensate is given by its minimum, which occurs
in the Landau gauge. Analyses in the effective potential formalism suggest
that a nonvanishing value of the dimension-two condensate is
favored~\cite{Verschelde:2001ia, Dudal:2011gd, Dudal:2019ing}; this is
supported by lattice QCD calculations of the gluon
propagator~\cite{Boucaud:2008gn}, the quark 
propagator~\cite{RuizArriola:2004en}, and also by 
a study based on resummation of Feynman diagrams~\cite{Dudal:2003vv}. 
As such a condensate leads to a dynamically generated gluon mass, it would
modify the infrared behavior of the gluon propagator. 
Remarkably, tree-level calculation in 
the RGZ theory leads to a good description of the lattice
measurement of the Landau-gauge gluon propagator~\cite{Dudal:2010tf, 
Cucchieri:2011ig}. 
This suggests that a perturbative calculation in the RGZ theory may be able to 
account for a good part of the nonperturbative dynamics of gluons in the
$SU(3)$ gauge theory. It would therefore be interesting to compute other
quantities involving gauge fields in the RGZ theory in perturbation theory, 
such as the QCD field-strength correlators and their moments. 

In this work, we compute the two-point correlation functions of the QCD
field-strength tensor and their moments in the RGZ theory at tree level. 
At tree level, perturbative calculations in the RGZ theory simply amounts to 
using the modified gluon propagator in usual perturbative QCD calculations. 
Due to the dimensionful parameters associated with the Gribov
parameter and the condensates, perturbative calculations in the RGZ theory are
infrared finite. When the field-strength correlators are computed
perturbatively in the RGZ theory, their long-distance behaviors are modified
from the power-law behaviors in the usual perturbative QCD calculations so that
their moments are infrared finite, while at short distances they reproduce
the leading UV behavior in perturbative QCD. 
We thus obtain finite values of the moments 
by subtracting the UV divergences through renormalization, 
and obtain renormalized color-octet matrix elements. 
From this we quantitatively determine the color-octet contributions to 
quarkonium decay rates, which can be compared directly with experiment.  

This paper is organized as follows. In Sec.~\ref{sec:QCDcorr}, 
we compute the two-point
correlation function of the QCD field-strength tensor in the RGZ theory, 
and compare the results with lattice QCD. 
In Sec.~\ref{sec:corr}, we compute the moments of the field-strength
correlation functions that appear in decay rates of heavy quarkonia. We use the
results for the
moments to obtain the color-octet matrix elements for quarkonium decays and
compute the color-octet contributions to decay rates of heavy quarkonia in
Sec.~\ref{sec:decays}. 
We conclude in Sec.~\ref{sec:conclusion}.

\section{QCD Field-Strength Correlators}
\label{sec:QCDcorr}

We define the two-point correlation function of the QCD field-strength tensor
in Euclidean space as 
\begin{equation}
\label{eq:corrdef}
D_{\mu \nu \rho \sigma} (x) = 
T_F \langle \Omega | 
g G_{\mu \nu}^a (x) \Phi_{ab} (x;0) 
g G_{\rho \sigma}^b (0) 
| \Omega \rangle,  
\end{equation}
where $| \Omega \rangle$ is the QCD vacuum, 
$G_{\mu \nu}^a=\partial_\mu A_\nu^a-\partial_\nu A_\mu^a
+g f^{abc} A_\mu^b A_\nu^c$ 
is the QCD field-strength tensor, 
$A_\mu$ is the gauge field, $g$ is the strong coupling, 
$T_F = 1/2$, 
and $\Phi_{ab} (x,0)$ is a straight Wilson line defined by 
\begin{equation}
\Phi(x,0)=P \exp \left[i g \int_0^1 d t \, x \cdot A^{\rm adj} (x t) \right], 
\end{equation}
where $P$ is the path ordering and $A^{\rm adj}$ is the gauge field in the
adjoint representation. 
The definition in Eq.~(\ref{eq:corrdef}) is consistent with 
Refs.~\cite{DiGiacomo:1992hhp, DElia:1997sdk, Bali:1997aj},
while it contains an additional factor of $g^2 T_F$ compared to what was used
in the perturbative calculation in Ref.~\cite{Eidemuller:1997bb}. 
The general form of the correlator can be written as 
\begin{align}
\label{eq:corrform}
D_{\mu \nu \rho \sigma} (x) &= 
( \delta_{\mu \rho} \delta_{\nu \sigma} - 
\delta_{\mu \sigma} \delta_{\nu \rho} )
\left[ {\cal D} (x^2) + {\cal D}_1 (x^2) \right] 
\nonumber \\ & \hspace{2.5ex}
+ ( \delta_{\mu \rho} x_\nu x_\sigma - \delta_{\mu \sigma} x_\nu x_\rho
\nonumber \\ & \hspace{6ex} 
- \delta_{\nu \rho} x_\mu x_\sigma + \delta_{\nu \sigma} x_\mu x_\rho )
\frac{\partial {\cal D}_1 (x^2)}{\partial x^2} ,
\end{align}
where ${\cal D} (x^2)$ and ${\cal D}_1 (x^2)$ are invariant functions of $x^2$. 
The form~(\ref{eq:corrform}) follows from the fact that ${\cal D} (x^2)$
vanishes in the Abelian gauge theory. Because of this, ${\cal D} (x^2)$
vanishes at tree level in the non-Abelian gauge theory. 

At tree level we can compute ${\cal D}_1 (x^2)$ at order $\alpha_s$ 
while ${\cal D} (x^2) = 0 + O(\alpha_s^2)$, where $\alpha_s =g^2/(4 \pi)$. 
Useful representations of the invariant functions can be obtained by defining
the two linear combinations 
\begin{align}
\label{eq:corrcombinations}
{\cal D}_E (x^2) &= \frac{1}{d-1} D_{0i0i} (x) 
\nonumber \\ &= 
{\cal D} (x^2) + {\cal D}_1 (x^2) + x^2 \frac{\partial {\cal D}_1
(x^2)}{\partial x^2}, 
\\
{\cal D}_B (x^2) &= \frac{1}{(d-1) (d-2)} D_{ijij} (x) 
\nonumber \\ &= 
{\cal D} (x^2) + {\cal D}_1 (x^2), 
\end{align}
where $d=4-2 \epsilon$ is the number of spacetime dimensions, 
and we work in a frame where
$x^\mu$ is aligned to the time $(\mu=0)$ direction. 
Note that in this frame, ${\cal D}_E(x^2)$ involves only the chromoelectric
fields and ${\cal D}_B (x^2)$ involves only the chromomagnetic fields. 
We work in arbitrary
spacetime dimensions to make possible the use of dimensional regularization in
the computation of the moments in the following section. 
The invariant functions can be conveniently computed in perturbation theory 
by using the first equalities. 
At tree level, the correlation function can be computed by differentiating the 
tree-level gluon propagator. The Landau-gauge gluon propagator in the RGZ
theory is given by~\cite{Dudal:2010tf}  
\begin{align}
\label{eq:RGZpropagator}
& 
\langle \Omega | A^a_\mu (x) A^b_\nu (y) | \Omega \rangle|_{\rm RGZ} 
=
\delta^{ab} 
\int \frac{d^dk}{(2 \pi)^d} e^{i k \cdot (x-y)} 
\nonumber \\ & \hspace{8ex}
\times 
\frac{
(\delta_{\mu \nu} - k_\mu k_\nu /k^2)
(k^2 +M^2)}{k^4 + (M^2+m^2)k^2+\lambda^4} + O(\alpha_s),
\end{align}
where $m$ and $M$ are dimension-one parameters associated with the condensates of
the gauge and auxiliary fields in the RGZ action, respectively, 
and $\lambda^4 = 2 g^2 N_c
\gamma^4 + m^2 M^2$ with $\gamma$ the Gribov parameter. 
Note that the tree-level propagator in the GZ theory is obtained by setting 
$m$ and $M$ to zero, while keeping $\gamma$ nonzero~\cite{Zwanziger:1989mf}. 
The values of $m$, $M$, and $\lambda$ that give a satisfactory description of 
the gluon propagator have been found in
Ref.~\cite{Dudal:2010tf} based on lattice QCD calculations of the gluon
propagator with coupling $\beta = 6/g^2$ set to $5.7$, $6.0$, and $6.2$. 
They read $M^2 = 2.15 \pm 0.13$~GeV$^2$, 
$m^2 = -1.81 \pm 0.14$~GeV$^2$, and $\lambda^4 = 0.26$~GeV$^4$. 
By using Eq.~(\ref{eq:RGZpropagator}) 
we obtain
\begin{align}
\label{eq:corrcombinationB_RGZ}
{\cal D}_B (x^2) |_{\rm RGZ} 
=& 
\frac{2 g^2 N_c C_F}{d-1} \int \frac{d^dk}{(2 \pi)^d} e^{i k_0 \sqrt{x^2}} 
\nonumber \\ & \times 
\frac{\bm{k}^2 (k^2 +M^2)}{k^4 + (M^2+m^2)k^2+\lambda^4} + O(\alpha_s^2),
\end{align}
where $C_F = (N_c^2-1)/(2 N_c)$. 
Note that this expression is valid in arbitrary dimensions. 
We can integrate over $k_0$ by using contour integration. The poles in $k_0$
are located at 
$\pm i \sqrt{\bm{k}^2 + (M^2+m^2 \pm Q^2)/2}$
and 
$\pm i \sqrt{\bm{k}^2 + (M^2+m^2 \mp Q^2)/2}$, 
where $Q^2 = i \sqrt{4 \lambda^4 - (M^2+m^2)^2}$. 
Note that $Q^2$ is purely imaginary 
for the values of the parameters determined in Ref.~\cite{Dudal:2010tf}. 
Because $\sqrt{x^2}>0$, we must close the contour on the upper half plane. 
We obtain
\begin{align}
\label{eq:corrcombinationB_RGZ2}
{\cal D}_B (x^2) |_{\rm RGZ}
=&
\frac{2 g^2 N_c C_F}{d-1}  
\int \frac{d^{d-1}k}{(2 \pi)^{d-1}} \bm{k}^2 
\nonumber \\ & \times 
\bigg[ 
\frac{e^{ - \kappa^+ \sqrt{x^2}} ( m^2-M^2+Q^2)} { 4 Q^2 \kappa^+}
\nonumber \\ & \hspace{3ex} 
+ 
\frac{e^{ - \kappa^- \sqrt{x^2}} ( m^2-M^2-Q^2)} { - 4 Q^2 \kappa^-} \bigg] 
\nonumber \\ & 
+ O(\alpha_s^2), 
\end{align}
where $\kappa^{\pm} = \sqrt{ \bm{k}^2 + (M^2+m^2 \pm Q^2)/2}$. 
Note that the quantity in the square brackets can be written as twice the
real part of the first term. 
To obtain ${\cal D}_B (x^2)$ at a given value of $x^2$, the remaining integral 
in $d-1=3$ dimensions can be computed numerically. 
Note that the small-$x^2$ behavior can be computed by setting 
the dimensionful parameters $m$, $M$, and $\lambda$ to zero, which leads to the
perturbative QCD result ${\cal D}_B (x^2) |_{\rm pQCD} = g^2 N_c C_F/(\pi^2
x^4)+ O(\alpha_s^2)$, in agreement with Ref.~\cite{Eidemuller:1997bb}. 

\begin{figure*}
\includegraphics[width=0.99\textwidth]{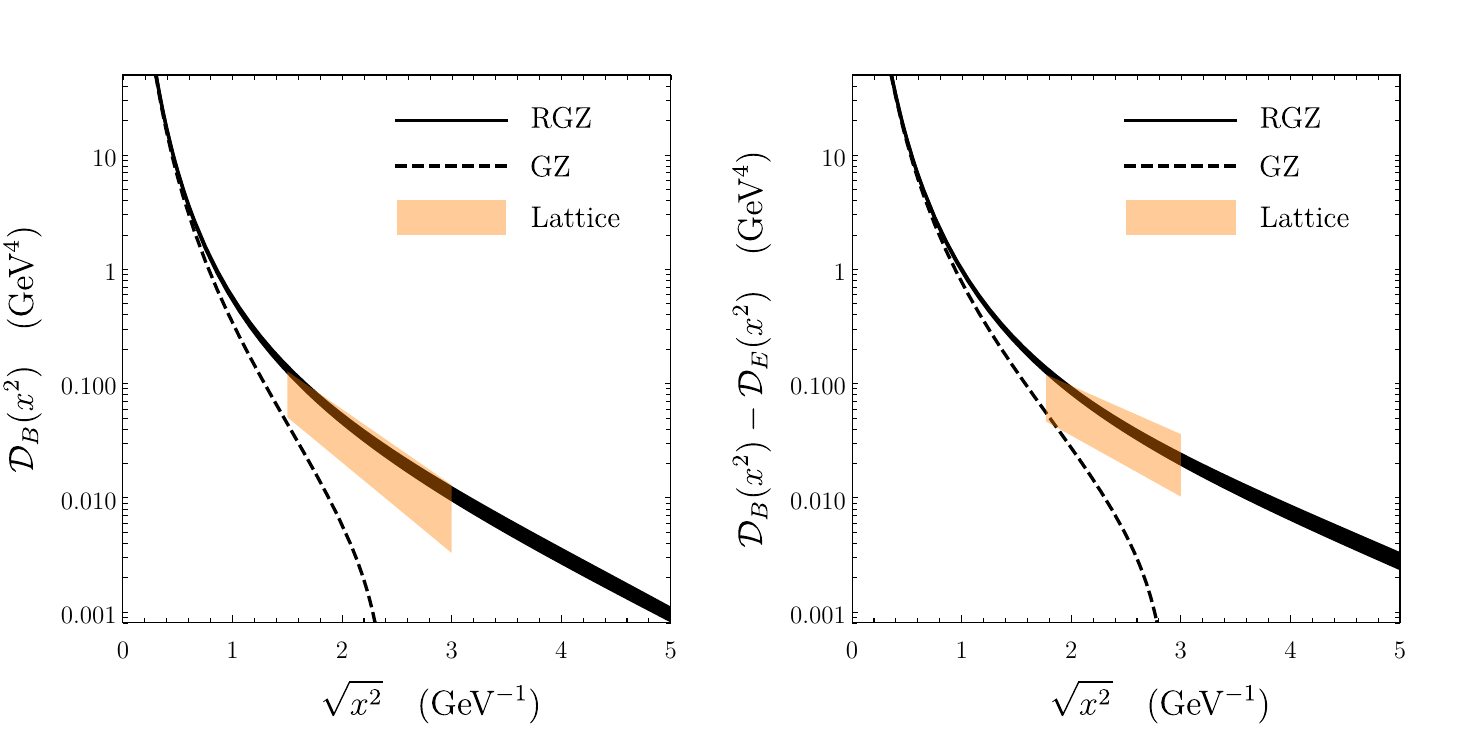}
\caption{\label{fig}
Results for the invariant functions ${\cal D}_B(x^2)$ (left panel) and 
${\cal D}_B(x^2) - {\cal D}_E(x^2)$ (right panel) from the 
RGZ theory (black bands) shown against 
results in the GZ theory (black dashed lines) and 
lattice QCD results from Ref.~\cite{Bali:1997aj} (orange bands). 
}
\end{figure*}

We show the numerical result for ${\cal D}_B(x^2)|_{\rm RGZ}$ in Fig.~\ref{fig} 
compared to the lattice measurement from Ref.~\cite{Bali:1997aj}. 
For the numerical calculation we set the strong coupling to be $\beta = 6.0$, 
which is
close to the average value used in the determination of the parameters of the 
RGZ theory in Ref.~\cite{Dudal:2010tf}, and is also same as the value used for 
the main results of the lattice study in Ref.~\cite{Bali:1997aj}.
Note that our definition of ${\cal D}_B(x^2)$ corresponds to $D_\perp (x^2)$ in
Ref.~\cite{Bali:1997aj}. 
We display the lattice measurement by using the functional form given by 
${D}_\perp^{\rm lat} (x^2) = A e^{-|x|/\lambda_a}$ 
with $A=0.94^{+0.32}_{-0.16}$~GeV$^4$ and $\lambda_a =
0.120^{+0.009}_{-0.012}$~fm determined in Ref.~\cite{Bali:1997aj}. 
Because this result was obtained from lattice data for $\sqrt{x^2} > 0.3$~fm,
and lattice data are only shown for $\sqrt{x^2} < 0.6$~fm in
Ref.~\cite{Bali:1997aj}, we only display the result from ${D}_\perp^{\rm lat}
(x^2)$ for $\sqrt{x^2}$ between $0.3$ and $0.6$~fm. 
We can see that the RGZ result agrees with the lattice QCD result within
uncertainties. 
Note that similarly to the lattice study, ${\cal D}_B(x^2)|_{\rm RGZ}$
exhibits an exponentially decaying behavior at large $x^2$ with a finite
correlation length, as can be seen from Fig.~\ref{fig} for $\sqrt{x^2}$ larger
than about 2~GeV$^{-1}$. 
A fit of the same functional form $C e^{ - |x|/\lambda_c}$ for $|x|$ 
larger than 2~GeV$^{-1}$ ($\approx 0.4$~fm) to the RGZ result leads to 
$C = 0.58 \pm 0.03$~GeV$^4$ and $\lambda_c = 0.77\pm0.01$~GeV$^{-1}$. 
This correlation length reads in distance units $\lambda_c \approx 0.15$~fm. 
The lattice QCD study in Ref.~\cite{Bali:1997aj} also found that at small
$\sqrt{x^2}$ the function ${\cal D}_B(x^2)$ shows a power-law behavior given by
$0.4/x^4$, which agrees very well with the tree-level perturbative QCD
result 
if we set the coupling using $\beta = 6.0$ as was done in the lattice study in
Ref.~\cite{Bali:1997aj}. For comparison, in Fig.~\ref{fig} we also show the
result from the GZ theory that we obtain by setting $m = M =0$ while keeping
the parameter $\gamma$ the same as the one obtained in the RGZ theory. 
We find that the GZ theory results in the long-distance behavior of 
${\cal D}_B(x^2)$ that falls off too quickly compared to the RGZ theory and is
incompatible with lattice QCD data. In fact, we find that the GZ theory result
for ${\cal D}_B(x^2)$ always falls off faster than the
perturbative QCD result for any nonzero value of the Gribov
parameter, so that it would not be possible to obtain a result that is
compatible with lattice QCD measurements. 

Now we look into ${\cal D}_E (x^2)$. Because ${\cal D} (x^2)$ vanishes at tree
level, we could obtain $x^2 \partial {\cal D}_1(x^2)/\partial x^2$ at tree
level by differentiating and multiplying by $x^2$ 
our result for ${\cal D}_B (x^2)$. 
However, for computing moments of the correlators in the following sections 
it is necessary to obtain a dimensionally regulated 
momentum-integral representation for ${\cal D}_E (x^2)$. 
By using the definition for ${\cal D}_E (x^2)$ and the tree-level gluon 
propagator in the RGZ theory we obtain 
\begin{align}
\label{eq:corrcombinationE_RGZ}
&
{\cal D}_E (x^2) |_{\rm RGZ}
= \frac{g^2 N_c C_F}{d-1} \int \frac{d^dk}{(2 \pi)^d} 
e^{i k_0 \sqrt{x^2}} 
\nonumber \\ & \quad \times
\frac{
[ \bm{k}^2 + (d-1) k_0^2 ]
(k^2 +M^2)}{k^4+(M^2+m^2) k^2 +\lambda^4} 
+ O(\alpha_s^2). 
\end{align}
We can again integrate over $k_0$ by using contour integration, closing the 
contour on the upper half plane. We obtain
\begin{align}
\label{eq:corrcombinationE_RGZ2}
& {\cal D}_E (x^2) |_{\rm RGZ}
=
\frac{g^2 N_c C_F}{d-1}
\int \frac{d^{d-1}k}{(2 \pi)^{d-1}} 
\nonumber \\ & 
\times \bigg[
\frac{\bm{k}^2 - (d-1) (\kappa^+)^2 }{4 Q^2}
\frac{e^{ - \kappa^+ \sqrt{x^2}} ( m^2-M^2+Q^2)} { \kappa^+}
\nonumber \\ & \quad
+
\frac{\bm{k}^2 - (d-1) (\kappa^-)^2 }{-4 Q^2} 
\frac{e^{ - \kappa^- \sqrt{x^2}} ( m^2-M^2-Q^2)} { \kappa^-}
\bigg]
\nonumber \\ & +O(\alpha_s^2) .
\end{align}
Similarly to the case of ${\cal D}_B (x^2)$, we can obtain 
${\cal D}_E (x^2)$ at a given
value of $x^2$ by evaluating the remaining integral in $d-1=3$ dimensions 
numerically. 

In order to compare with lattice data, we consider the combination
${\cal D}_B (x^2) - {\cal D}_E (x^2) = 
- x^2 \partial {\cal D}_1(x^2)/\partial x^2$
which corresponds to $-D_* (x^2)$ in the 
lattice QCD study in Ref.~\cite{Bali:1997aj}. 
We show the result in Fig.~\ref{fig}
against the lattice measurement from Ref.~\cite{Bali:1997aj}.
We display the lattice measurement by using the functional form given by
$- {D}_*^{\rm lat} (x^2) = B e^{-|x|/\lambda_b}$
with $B=0.47^{+0.20}_{-0.06}$~GeV$^4$ and $\lambda_b =
0.189^{+0.013}_{-0.029}$~fm in Ref.~\cite{Bali:1997aj}.
Because the fit in Ref.~\cite{Bali:1997aj} for ${D}_*^{\rm lat} (x^2)$ was 
restricted to $\sqrt{x^2} > 0.35$~fm, 
and lattice data are only shown for $\sqrt{x^2} < 0.6$~fm, 
we only display the lattice result 
for $\sqrt{x^2}$ between $0.35$ and $0.6$~fm.
Similarly to the case of ${\cal D}_B (x^2)$, the result for 
${\cal D}_B (x^2) - {\cal D}_E (x^2)$
in the RGZ theory agrees 
with the lattice result in Ref.~\cite{Bali:1997aj} within uncertainties. 
The authors of Ref.~\cite{Bali:1997aj} did not present a functional form of the 
short-distance behavior of ${\cal D}_B (x^2) - {\cal D}_E (x^2)$. 
Nevertheless, the lattice QCD data available in Ref.~\cite{Bali:1997aj} 
at distances shorter than about 0.35~fm 
seem to agree very well with what is expected from
perturbative QCD at tree level given by 
${\cal D}_B (x^2) - {\cal D}_E (x^2) |_{\rm pQCD} \approx 0.8/x^4$ for $\beta =
6.0$. 
As we have done for ${\cal D}_B (x^2)$, we also show the results from the GZ
theory in Fig.~\ref{fig}, which we obtain by setting $m=M=0$ and keeping the
Gribov parameter $\gamma$ the same as the one obtained in the RGZ theory. 
We see that also in the case of ${\cal D}_B (x^2) - {\cal D}_E (x^2)$
the result from the GZ theory has a long-distance behavior that falls off too
quickly and is incompatible with lattice QCD results. 

As we have discussed earlier, in perturbation theory ${\cal D}(x^2)$ appears
from next-to-leading order accuracy, and so, at tree level only the ${\cal
D}_1(x^2)$ and its derivative contribute to ${\cal D}_B (x^2)$ and 
${\cal D}_E (x^2)$. Based on the fact that the tree-level results for the
invariant functions in the RGZ theory are compatible with the lattice QCD study
in Ref.~\cite{Bali:1997aj}, we may assume that ${\cal D}(x^2)$ is indeed small
and is negligible compared to the uncertainties in the lattice results. 
The agreement in the short-distance behaviors of the invariant functions
between the lattice QCD study and the RGZ results supports the 
suppression of ${\cal D}(x^2)$ even at short distances. 
We may estimate the effects from a
nonzero ${\cal D}(x^2)$ by using the lattice QCD results in
Ref.~\cite{Bali:1997aj} in our numerical analysis. 

We note that there are other lattice QCD studies of the two-point
field-strength correlation function based on the cooling method in
Refs.~\cite{DiGiacomo:1992hhp, DElia:1997sdk}. 
These lattice studies result in a much larger 
${\cal D}(x^2)$ compared to ${\cal D}_1(x^2)$ at both short and long distances, 
which contradicts the na\"ive expectation from perturbation theory that 
${\cal D}(x^2)$ would be suppressed compared to ${\cal D}_1(x^2)$, 
and is also in conflict with the smallness of ${\cal D}(x^2)$ that can be
inferred from the lattice study in Ref.~\cite{Bali:1997aj}.
As was discussed in Ref.~\cite{Eidemuller:1997bb}, the two-point field-strength
correlation function involves UV divergences at loop level, 
and after renormalization the functions ${\cal D}(x^2)$ and ${\cal D}_1(x^2)$
mix under change of the renormalization scheme and scale; since the cooling
method used in Refs.~\cite{DiGiacomo:1992hhp, DElia:1997sdk} removes
short-range fluctuations, it would not be possible to make direct comparisons
until the results are converted to a common scheme. 
In our case, because we work at tree level and the lattice results in
Ref.~\cite{Bali:1997aj} are renormalized nonperturbatively, it is not possible
to quantify the effect of scheme dependence; nonetheless, the good agreement
between the tree-level RGZ results and the lattice results in
Ref.~\cite{Bali:1997aj} at both long and short distances may indicate that the
effect of the scheme change is small and may even be negligible between the two
results. 

So far the calculation in this section has been done in the Landau gauge. 
Because we work with the gauge-invariant definition of the
field-strength correlator~(\ref{eq:corrdef}), we expect that the results for 
${\cal D}_E$ and ${\cal D}_B$ we obtain to be valid in any gauge. We may
check this by, for example, computing the correlators from the gluon
propagator in a covariant gauge. 
We first need to replace the dimension-two condensate term $A^2$ in the action
with a gauge-invariant expression~\cite{Zwanziger:1990tn, Lavelle:1995ty}
\begin{equation}
\label{eq:Asqinvariant}
A_\mu^a \left( \delta_{\mu \nu} - \frac{\partial_\mu \partial_\nu}{\partial^2}
\right) A_\nu^a + O(g),
\end{equation}
where the corrections from order $g$ involve products of three or more gauge
fields and can be computed systematically as formal power series in $A$. 
At the current level of accuracy, we only need to keep the lowest-order 
nonvanishing contribution. 
We can then obtain the gluon propagator in the RGZ theory in a generic 
covariant gauge; the result in momentum space contains an extra term 
$\alpha k^2/[k^4+2 \alpha  g^2 N_c \gamma^4 k^2/(k^2+M^2)] \times 
k_\mu k_\nu/k^2$ compared to the Landau-gauge expression 
in Eq.~(\ref{eq:RGZpropagator}), with $\alpha$ the gauge parameter. 
It is easy to check explicitly that this additional term makes vanishing 
contributions in ${\cal D}_E$ and ${\cal D}_B$. This may be understood as a 
consequence of the Ward identity, arising from the gauge invariance of the 
field-strength correlator. 

In this section we have established the two-point field-strength correlation
functions at tree level in the RGZ theory as dimensionally regulated momentum
integrals in Eqs.~(\ref{eq:corrcombinationB_RGZ2}) and
(\ref{eq:corrcombinationE_RGZ2}). 
We will use these results in the following sections to compute moments of the
correlation functions in dimensional regularization.

\section{Moments of field-strength correlators} 
\label{sec:corr}

In this section we compute the moments of the two-point field-strength
correlator. Rather than considering all possible moments of the correlator, we
focus on the ones that appear in heavy quarkonium decay rates. 
Roughly speaking, the moments of the correlator represent the squared amplitude
for a heavy quark-antiquark pair in a color-octet state to evolve into a
color-singlet state through insertions of chromoelectric or chromomagnetic
fields on the heavy quark or antiquark lines. 
We refer the readers to Refs.~\cite{Brambilla:2001xy, Brambilla:2002nu,
Brambilla:2020xod} for details of the pNRQCD formalism that were used to obtain
expressions for color-octet NRQCD matrix elements in terms of moments of
field-strength correlators. 

The following dimensionless moment of ${\cal D}_E (x^2)$ defined by 
\begin{equation}
\label{eq:E3def}
{\cal E}_3 = - \frac{d-1}{N_c} 
\int_0^\infty d \tau \, \tau^3 {\cal D}_E (\tau^2), 
\end{equation}
appears in inclusive decay rates of $P$-wave heavy quarkonia at leading order
in $v$. The factor $1/N_c$ appears from the projection of the $Q \bar Q$ color,
and the sign arises from the fact that ${\cal E}_3$ was defined 
in Minkowski space in Ref.~\cite{Brambilla:2002nu} while the right-hand side is
computed in Euclidean space. 
Because the ${\cal D}_E (\tau)$ scales like $1/\tau^4$ at small $\tau$,
the integral over $\tau$ contains a logarithmic UV divergence. After
renormalization, ${\cal E}_3$ acquires a scale dependence given by 
\begin{equation}
\label{eq:E3RG}
\frac{d}{d \log \Lambda}
{\cal E}_3^{(\Lambda)} = \frac{12 \alpha_s C_F}{\pi} + O(\alpha_s^2), 
\end{equation}
where $\Lambda$ is the renormalization scale for ${\cal E}_3$. 
Throughout this paper, we use the superscript $(\Lambda)$ to denote that the
quantity is renormalized in the $\overline{\rm MS}$ scheme and depends on the
$\overline{\rm MS}$ scale $\Lambda$. 
It has been known that this scale dependence directly reproduces the
renormalization-scale dependence of the color-octet matrix element that appears
in the $P$-wave quarkonium decay rates~\cite{Bodwin:1994jh, Brambilla:2001xy}. 
Note that ${\cal E}_3$ also appears in the order-$v^2$ correction to the
leading-order color-singlet matrix elements for $S$-wave quarkonium
decays~\cite{Brambilla:2002nu}. 

Similarly, the following dimension-two moments of ${\cal D}_E (x^2)$ and 
${\cal D}_B (x^2)$ defined by 
\begin{subequations}
\label{eq:E1B1defs}
\begin{align}
{\cal E}_1 &= \frac{d-1}{N_c} \int_0^\infty d \tau\, \tau {\cal D}_E (\tau^2), 
\\
{\cal B}_1 &= \frac{(d-1) (d-2)}{2 N_c} \int_0^\infty d \tau\, \tau 
{\cal D}_B (\tau^2), 
\end{align}
\end{subequations}
appear in inclusive decay rates of $S$-wave heavy quarkonia at relative order
$v^4$ and $v^3$, respectively~\cite{Brambilla:2002nu}. 
Although the contributions from these moments to the decay rates are suppressed
by powers of $v$, their effects can be enhanced by large short-distance 
coefficients. In the case of spin-1 $S$-wave heavy quarkonium decays, 
the short-distance coefficients associated with color-octet contributions are
enhanced by $1/\alpha_s$ compared to the one at leading order in $v$, which can
make the color-octet contributions numerically
significant~\cite{Petrelli:1997ge, Bodwin:2002cfe}.
In the decays of spin-0 $S$-wave states, such as the $\eta_c$ and $\eta_b$, 
the lack of knowledge of the color-octet matrix element
arising from ${\cal B}_1$ is a significant source of 
uncertainties~\cite{Bodwin:2001pt, Brambilla:2018tyu}. 
The quantity ${\cal E}_1$ also appears in the order-$v^2$ correction to
two-photon decay rates of $P$-wave heavy quarkonia~\cite{Brambilla:2002nu,
Brambilla:2020xod}. 
Hence, accurate determinations of ${\cal E}_1$ and ${\cal B}_1$ are
phenomenologically important. 
Note that both ${\cal E}_1$ and ${\cal B}_1$ contain power UV divergences at
small $\tau$, which must be subtracted in dimensional regularization
consistently with the calculation of short-distance coefficients in
perturbation theory. 

We also compute the dimension-one moment $i {\cal E}_2$ defined by 
\begin{equation}
i {\cal E}_2 = \frac{d-1}{N_c} \int_0^\infty d \tau\, \tau^2 
{\cal D}_E (\tau^2), 
\end{equation}
where the phase is chosen to recover the Minkowski space definition in
Ref.~\cite{Brambilla:2002nu}. Note that ${\cal E}_2$ is purely imaginary, so
that $i {\cal E}_2$ is real. Although this moment does not appear directly in
color-octet matrix elements, it appears in the corrections to the quarkonium
wave functions at the origin associated with the velocity-dependent potential at
zero separation~\cite{Chung:2020zqc, Chung:2021efj}, and can be useful in heavy
quarkonium decay phenomenology. 

We note that the overall phases of the right-hand sides of the definitions of
the moments can be checked against the Minkowski space definitions in
Ref.~\cite{Brambilla:2002nu} by computing them in perturbative QCD and
comparing the integrands of the momentum integral. 

We now proceed to compute the moments in the RGZ theory. 

\subsection{\boldmath Dimensionless moment ${\cal E}_3$}

We first compute the dimensionless moment ${\cal E}_3$. 
By using Eq.~(\ref{eq:corrcombinationE_RGZ2}) 
and the identity $\int_0^\infty d \tau\, \tau^3 \exp(- A \tau) = 6/A^4$,
which is valid when ${\rm Re} A > 0$, 
we obtain the following expression
\begin{align}
\label{eq:E3integral}
{\cal E}_3 |_{\rm RGZ} 
&=
-
6 g^2 C_F
\int \frac{d^{d-1}k}{(2 \pi)^{d-1}}
\nonumber \\ & \hspace{3ex} \times 
\bigg[
\frac{\bm{k}^2 - (d-1) (\kappa^+)^2 }{4 Q^2}
\frac{m^2-M^2+Q^2} {( \kappa^+)^5}
\nonumber \\ & \hspace{7ex} 
+
\frac{\bm{k}^2 - (d-1) (\kappa^-)^2 }{-4 Q^2}
\frac{m^2-M^2-Q^2} {(\kappa^-)^5}
\bigg] 
\nonumber \\ & \hspace{3ex} 
+ O(\alpha_s^2),
\end{align}
which is valid in arbitrary dimensions. 
The $d-1$-dimensional integral over $\bm{k}$ is straightforwardly evaluated 
by using 
\begin{equation}
\int \frac{d^{d-1}k}{(2 \pi)^{d-1}} 
= \frac{2 \pi^{(d-1)/2}\mu^{4-d}
}{(2 \pi)^{d-1} \Gamma (\tfrac{d-1}{2})} 
\int_0^\infty d |\bm{k}| \, |\bm{k}|^{d-2} , 
\end{equation}
where $\mu$ is a scale associated with dimensional regularization. 
We obtain 
\begin{align}
\label{eq:E3integral_result}
&
{\cal E}_3 |_{\rm RGZ}
=
-
6 g^2 C_F
\mu^{2 \epsilon} 
\frac{(-3+2 \epsilon) \Gamma(\epsilon) }
{3 \times 8^{1-\epsilon} \pi^{2-\epsilon} Q^2}
\nonumber \\ & \hspace{3ex} 
\times \bigg[ 
(m^2 +M^2+Q^2)^{-\epsilon} (m^2 -M^2+Q^2)
\nonumber \\ & \hspace{6ex} 
+ 
(m^2 +M^2-Q^2)^{-\epsilon} (M^2 -m^2+Q^2)\bigg]
+O(\alpha_s^2) 
\nonumber \\
&= 
\frac{3 g^2 C_F}{2 \pi^2} \left[ \frac{1}{\epsilon} 
+ \log (2 \Lambda^2/M^2) - \frac{2}{3} \right]
\nonumber \\ & \hspace{3ex} 
+ \frac{3 g^2 C_F}{4 \pi^2 Q^2} 
\bigg[ (m^2-M^2-Q^2) \log \left(\frac{m^2+M^2-Q^2}{M^2} \right)
\nonumber \\ & \hspace{12ex} 
- (m^2-M^2+Q^2) \log \left(\frac{m^2+M^2+Q^2}{M^2} \right) \bigg]
\nonumber \\ & \hspace{3ex} 
+ O(\alpha_s^2, \epsilon) , 
\end{align}
where in the last equality we expanded in powers of $\epsilon =(4-d)/2$ 
and set 
$\mu^2 = \Lambda^2 e^{\gamma_{\rm E}}/(4 \pi)$, so that $\Lambda$ is a
$\overline{\rm MS}$ scale. 
Here, $\gamma_{\rm E}$ is the Euler-Mascheroni constant. 
Note that the result in the last equality is invariant under simultaneous 
rescalings of the denominator factors $M^2$ in the arguments of the
logarithms. 
The pole in $\epsilon$ is exactly what we expect from the order-$\alpha_s$ 
scale dependence in Eq.~(\ref{eq:E3RG}).
After renormalization we then have in the $\overline{\rm MS}$ scheme 
\begin{align}
\label{eq:E3integral_renorm}
& 
{\cal E}_3^{(\Lambda)} |_{\rm RGZ} 
=
\frac{3 g^2 C_F}{2 \pi^2} \left[ 
\log (2 \Lambda^2/M^2) - \frac{2}{3} \right]
\nonumber \\ & \hspace{3ex}
+ \frac{3 g^2 C_F}{4 \pi^2 Q^2}
\bigg[ (m^2-M^2-Q^2) \log \left(\frac{m^2+M^2-Q^2}{M^2} \right)
\nonumber \\ & \hspace{12ex}
- (m^2-M^2+Q^2) \log \left(\frac{m^2+M^2+Q^2}{M^2} \right) \bigg]
\nonumber \\ & \hspace{3ex}
+ O(\alpha_s^2),
\end{align}
with $\Lambda$ the $\overline{\rm MS}$ renormalization scale. 
This is our result for ${\cal E}_3$ in the RGZ theory at tree level. 
Note that the explicit dependence on $\log \Lambda$ satisfies the
evolution equation in Eq.~(\ref{eq:E3RG}).

\subsection{\boldmath Dimension-two moments ${\cal E}_1$ and ${\cal B}_1$}

Now we consider the dimension-two moments 
${\cal E}_1$ and ${\cal B}_1$.  We first compute ${\cal E}_1$.  
From Eq.~(\ref{eq:corrcombinationE_RGZ2}) and the identity 
$\int_0^\infty d \tau \, \tau \exp(-A \tau) = 1/A^2$ we obtain 
\begin{align}
\label{eq:E1integral}
{\cal E}_1 |_{\rm RGZ}
&=
g^2 C_F
\int \frac{d^{d-1}k}{(2 \pi)^{d-1}}
\nonumber \\ & \quad \times 
\bigg[
\frac{\bm{k}^2 - (d-1) (\kappa^+)^2 }{4 Q^2}
\frac{m^2-M^2+Q^2} {( \kappa^+)^3}
\nonumber \\ & \quad \quad
+
\frac{\bm{k}^2 - (d-1) (\kappa^-)^2 }{-4 Q^2}
\frac{m^2-M^2-Q^2} {(\kappa^-)^3}
\bigg] 
\nonumber\\ & \quad
+ O(\alpha_s^2),
\end{align}
which is valid for arbitrary $d$. The ($d-1$)-dimensional integral gives 
\begin{align}
\label{eq:E1integralcompute}
{\cal E}_1 |_{\rm RGZ}
&=
\frac{g^2 C_F \mu^{2 \epsilon} }{(8 \pi)^{2- \epsilon} Q^2}
(3- 2\epsilon) \Gamma(\epsilon-1) 
\nonumber \\ & \quad \times 
\bigg\{ 
\bigg[ ( m^2+M^2-Q^2)^{1-\epsilon} (M^2-m^2+Q^2) 
\nonumber \\ & \hspace{6ex} 
+ (m^2+M^2+Q^2)^{1-\epsilon} (m^2-M^2+Q^2) \bigg] 
\nonumber \\ & \hspace{5ex}
- \bigg[  ( m^2+M^2-Q^2)^{1-\epsilon} (M^2-m^2+Q^2)
\nonumber \\ & \hspace{6ex}
+ (m^2+M^2+Q^2)^{1-\epsilon} (m^2-M^2+Q^2) \bigg]
\bigg\} 
\nonumber \\ & \quad
+ O(\alpha_s^2), 
\end{align}
where the terms in the first square brackets come from the part of the 
integrand in Eq.~(\ref{eq:E1integral}) that is proportional to $\bm{k}^2$, 
while the terms in the second square brackets 
come from the part that is proportional to $d-1$. 
Note the appearance of the pole at $d = 2$ from $\Gamma(\epsilon-1)$, 
indicating that the individual integrals are quadratically power divergent.
However, the contributions from the two parts of the integrand cancel each
other, so that ${\cal E}_1|_{\rm RGZ}$ vanishes through order $g^2$:
\begin{equation}
{\cal E}_1|_{\rm RGZ} = 0 + O(\alpha_s^2), 
\end{equation}
which is valid for arbitrary $d$. 

We can understand this result for ${\cal E}_1$ in terms of the functions 
${\cal D}(x^2)$ and ${\cal D}_1(x^2)$ by rewriting the definition as 
\begin{align}
\label{eq:E1reexpress}
{\cal E}_1
&=
\frac{d-1}{2 N_c}
\int_0^\infty d \tau^2 \left( {\cal D} (\tau^2) + {\cal D}_1 (\tau^2) + 
\tau^2 \frac{\partial {\cal D}_1 (\tau^2)}{\partial \tau^2} \right)
\nonumber\\
&=
\frac{d-1}{2 N_c}
\bigg[
\left( \int_0^\infty d \tau^2 {\cal D} (\tau^2) \right) 
+ 
\tau^2 {\cal D}_1 (\tau^2) \Big|_{\tau^2\to \infty}
\nonumber\\ & \hspace{10ex} 
-
\tau^2 {\cal D}_1 (\tau^2) \Big|_{\tau^2=0}
\bigg],
\end{align}
where we changed the integration variable to $\tau^2$, 
and used integration by parts 
to obtain the last equality. Note that this expression is valid to all orders. 
While $\tau^2 {\cal D}_1 (\tau^2)|_{\tau^2 = \infty} = 0$, 
the vanishing of the last term in the last equality at $\tau^2=0$ follows from
the fact that the the UV divergences of ${\cal E}_1$ must be regularized
dimensionally. 
For example, in dimensional regularization the short-distance behavior of 
${\cal D}_1 (\tau^2)$ is proportional to $1/\tau^d$, which must be computed
with $d<2$ in order to regulate power UV divergences; hence we obtain 
$\tau^2 {\cal D}_1 (\tau^2)|_{\tau^2 = 0} = 0$. 
Therefore, in dimensional regularization ${\cal E}_1$ comes solely from 
${\cal D}(\tau^2)$, and because this appears only from order $\alpha_s^2$,
${\cal E}_1$ vanishes at tree level. 
Note that this result is valid not just in the RGZ theory, but generally holds
in a non-Abelian gauge theory. 

Now we compute ${\cal B}_1$. We use Eq.~(\ref{eq:corrcombinationB_RGZ2}) and 
the identity $\int_0^\infty d \tau \, \tau \exp(-A \tau) = 1/A^2$ to obtain 
\begin{align}
\label{eq:B1integral}
{\cal B}_1 |_{\rm RGZ}
&=
(d-2) g^2 C_F 
\int \frac{d^{d-1}k}{(2 \pi)^{d-1}}
\bm{k}^2 
\nonumber \\ & \quad \times 
\bigg[
\frac{m^2-M^2+Q^2} {4 Q^2 (\kappa^+)^3}
+
\frac{m^2-M^2-Q^2} {- 4 Q^2 (\kappa^-)^3}
\bigg] 
\nonumber \\ & \quad 
+ O(\alpha_s^2),
\end{align}
which is valid in arbitrary dimensions. A straightforward evaluation of the
($d-1$)-dimensional integral over $\bm{k}$ yields 
\begin{align}
\label{eq:B1integral_result}
{\cal B}_1 |_{\rm RGZ}
&=
(d-2) g^2 C_F 
\mu^{2 \epsilon}
\frac{(3-2 \epsilon) \Gamma(\epsilon-1) }
{(8 \pi)^{2-\epsilon} Q^2}
\nonumber \\ & \quad
\times \bigg[
(m^2 +M^2-Q^2)^{1-\epsilon} (M^2 -m^2+Q^2)
\nonumber \\ & \quad \quad 
+
(m^2 +M^2+Q^2)^{1-\epsilon} (m^2 -M^2+Q^2)\bigg]
\nonumber \\ & \quad 
+O(\alpha_s^2). 
\end{align}
Note the appearance of a power divergence at $d=2$. 
The result also contains a logarithmic UV divergence; if we expand in powers of
$\epsilon$, we obtain 
\begin{align}
\label{eq:B1integral_result2}
& {\cal B}_1 |_{\rm RGZ}
=
- 
g^2 C_F
\frac{3m^2}{8 \pi ^2} \left[
\frac{1}{\epsilon} +
\log (2 \Lambda^2/M^2) - \frac{2}{3}\right]
\nonumber \\ & \quad 
-\frac{3 g^2 C_F}{32 \pi ^2 Q^2}
\bigg\{
\left[\left(m^2-Q^2\right)^2-M^4\right] 
\log \left(\frac{m^2+M^2-Q^2}{M^2}\right)
\nonumber \\ & \quad \quad
-\left[\left(m^2+Q^2\right)^2-M^4\right] 
\log \left(\frac{m^2+M^2+Q^2}{M^2}\right) 
\bigg\}
\nonumber \\ & 
+ O(\alpha_s^2, \epsilon). 
\end{align}
Note that the UV pole is proportional to $m^2$. 
Since $m^2$ is the parameter associated with the dimension-two condensate of
the gauge field, this UV pole is of nonperturbative origin and does not have a
counterpart in the NRQCD factorization formalism.
Hence, it is not possible to properly renormalize this UV divergence 
within the NRQCD formalism. 
However, this logarithmic UV divergence can be related to the dimension-two
condensate at tree level, which contains the same divergence. 
In the Landau gauge the dimension-two condensate 
$\langle (g A)^2 \rangle \equiv \langle \Omega | g^2 A^a_\mu A^a_\mu | \Omega
\rangle$ can be computed at tree level as 
\begin{align}
\label{eq:AA}
&
\langle (g A)^2 \rangle|_{\rm RGZ}
\nonumber \\ &=
g^2 
\int \frac{d^dk}{(2 \pi)^d} 
\frac{(N_c^2-1) 
(d - 1)
(k^2 +M^2)}{k^4 + (M^2+m^2)k^2+\lambda^4} 
+ O(\alpha_s^2)
\nonumber \\ &=
g^2 (N_c^2-1)  (d-1)
\nonumber \\ & \quad \times 
\int \frac{d^{d-1}k}{(2 \pi)^{d-1}}
\bigg( \frac{m^2-M^2+Q^2} {4 Q^2 \kappa^+} +
\frac{m^2-M^2-Q^2} {- 4 Q^2 \kappa^-} \bigg) 
\nonumber \\ & \quad 
+ O(\alpha_s^2).
\end{align}
The first equality follows from the Landau-gauge gluon propagator at zero
separation, and the second equality is obtained by integrating over $k_0$ using
contour integration. 
Note that, in a general gauge, $A^a_\mu A^a_\mu$ must be replaced by a
gauge-invariant expression given in Eq.~(\ref{eq:Asqinvariant}). In this case,
any additional term in the gluon propagator involving $k_\mu$ or $k_\nu$ makes
a vanishing contribution to Eq.~(\ref{eq:AA}), so that the result is gauge
invariant. 
We compute the remaining integral to obtain
\begin{align}
\label{eq:AAresult}
&
\langle (g A)^2 \rangle|_{\rm RGZ}
= 
g^2 (N_c^2-1) 
\mu^{2 \epsilon}
\frac{(3-2 \epsilon) \Gamma(\epsilon-1) }
{(8 \pi)^{2-\epsilon} Q^2}
\nonumber \\ & \quad 
\times \bigg[
(m^2 +M^2-Q^2)^{1-\epsilon} (M^2 -m^2+Q^2)
\nonumber \\ & \quad \quad 
+
(m^2 +M^2+Q^2)^{1-\epsilon} (m^2 -M^2+Q^2)\bigg]
\nonumber \\ & \quad 
+O(\alpha_s^2).
\end{align}
By comparing this result with Eq.~(\ref{eq:B1integral_result}) we find that 
\begin{equation}
\label{eq:B1result_AA}
{\cal B}_1 |_{\rm RGZ}
= \frac{T_F}{N_c} (d-2) 
\langle (g A)^2 \rangle|_{\rm RGZ}
+ O(\alpha_s^2).
\end{equation}
Note that this relation holds for arbitrary $d$. 
The relation in Eq.~(\ref{eq:B1result_AA}) arises in a similar fashion 
as the vanishing of ${\cal E}_1$ at tree level; note that the integrand of 
${\cal B}_1$ in Eq.~(\ref{eq:B1integral}) is identical to the term proportional
to $\bm{k}^2$ in the integrand of ${\cal E}_1$ in Eq.~(\ref{eq:E1integral}),
and the integrand of $\langle (g A)^2 \rangle$ in Eq.~(\ref{eq:AA}) is just the
term proportional to $d-1$ in the integrand of ${\cal E}_1$ in
Eq.~(\ref{eq:E1integral}). As we have seen from the evaluation of ${\cal E}_1$,
the two contributions are equal, leading to the relation in
Eq.~(\ref{eq:B1result_AA}). Similarly to the case of ${\cal E}_1$, this
relation may be modified at higher orders in $\alpha_s$, notably from a nonzero
${\cal D} (\tau^2)$. 

In fact, the relation in Eq.~(\ref{eq:B1result_AA}) can be understood in terms
of gauge fields by rewriting the condensate in a manifestly gauge-invariant 
form in terms of the field-strength tensor, given by~\cite{Zwanziger:1990tn} 
\begin{equation}
\label{eq:AArelation_fieldstrength}
\langle A_\mu^a A_\mu^a\rangle|_{\textrm{Landau gauge}} 
= -\frac{1}{2} \langle G_{\mu \nu}^a (D^2)^{-1} G_{\mu \nu}^a \rangle + O(g),
\end{equation}
where $D$ is the covariant derivative. 
The corrections from order $g$ involve products of three or more
field-strength tensors. Then, by using 
$\langle G_{\mu \nu}^a (D^2)^{-1} G_{\mu \nu}^a \rangle
= (d-2)^{-1} \times \langle G_{\mu \nu}^a (D_0^2)^{-1} G_{\mu \nu}^a \rangle$, 
which follows from rotational invariance and taking a $d$-dimensional angular 
average, 
and by exponentiating $(D_0^2)^{-1}$ in momentum space at order $g^0$, we
obtain 
\begin{align}
\label{eq:AArelation_fieldstrength_v2}
\langle (g A)^2 \rangle
&= \frac{1}{4 T_F (d-2)} \int_0^\infty d x^2 
D_{\mu \nu \mu \nu} (x) + O(\alpha_s^2)
\nonumber \\
&= \frac{N_c}{T_F (d-2)} 
\left( {\cal E}_1 + {\cal B}_1 \right)
+ O(\alpha_s^2). 
\end{align}
This, together with the vanishing of ${\cal E}_1$ at order $g^2$, leads to the 
the relation in Eq.~(\ref{eq:B1result_AA}). 

As can be obtained from Eq.~(\ref{eq:AAresult}), or from the relation in
Eq.~(\ref{eq:B1result_AA}) and the result for ${\cal B}_1$ in
Eq.~(\ref{eq:B1integral_result2}), the UV pole in $\langle (g A)^2 \rangle$
that we obtain reproduces the lowest-order result in Eq.~(22) of
Ref.~\cite{Verschelde:2001ia} for the counterterm of the generating functional 
proportional to $m^2$. 
While perturbative calculations of ${\cal B}_1$ and $\langle (g A)^2 \rangle$ 
both yield logarithmic UV divergences, a finite result for the 
$\langle (g A)^2 \rangle$ has been obtained in the effective potential
approach~\cite{Verschelde:2001ia}:
\begin{equation}
\label{eq:AAvalue}
\langle (g A)^2 \rangle = - \frac{9}{13} \frac{N_c^2-1}{N_c} m^2. 
\end{equation}
This result is obtained by minimizing the effective potential in the presence
of the dimension-two condensate. 
Since this finite result has been obtained after subtracting the $1/\epsilon$
pole, the $\epsilon$-dependent coefficient in Eq.~(\ref{eq:B1result_AA})
produces an additional finite piece. That is, 
\begin{equation}
\label{eq:B1result_AA_eps}
{\cal B}_1 |_{\rm RGZ}
= \frac{2 T_F}{N_c} 
\langle (gA)^2 \rangle|_{\rm RGZ}
+ \frac{3 g^2 C_F m^2}{8 \pi^2} 
+ O(\alpha_s^2, \epsilon).
\end{equation}
From this and Eq.~(\ref{eq:AAvalue}) we obtain
\begin{equation}
\label{eq:B1result_AAfinite}
{\cal B}_1 |_{\rm RGZ} = - \frac{18 C_F}{13 N_c} m^2 
+ \frac{3 g^2 C_F m^2}{8 \pi^2} 
+ O(\alpha_s^2).
\end{equation}
Note that since $m^2$ is negative, the first term is positive while the second
term is negative. 
This is our finite result for ${\cal B}_1$ in the RGZ theory. 

\subsection{\boldmath Dimension-one moment $i {\cal E}_2$}

From Eq.~(\ref{eq:corrcombinationE_RGZ2}) and the identity
$\int_0^\infty d \tau \, \tau^2 \exp(-A \tau) = 2/A^3$ we obtain
\begin{align}
\label{eq:E2integral}
i {\cal E}_2 |_{\rm RGZ}
&=
2 g^2 C_F
\int \frac{d^{d-1}k}{(2 \pi)^{d-1}}
\nonumber \\ & \quad \times 
\bigg[
\frac{\bm{k}^2 - (d-1) (\kappa^+)^2 }{4 Q^2}
\frac{m^2-M^2+Q^2} {( \kappa^+)^4}
\nonumber \\ & \quad \quad 
+
\frac{\bm{k}^2 - (d-1) (\kappa^-)^2 }{-4 Q^2}
\frac{m^2-M^2-Q^2} {(\kappa^-)^4} \bigg] 
\nonumber \\ & \quad 
+ O(\alpha_s^2).
\end{align}
The integral can be evaluated straightforwardly:
\begin{align}
\label{eq:E2integral2}
&
i {\cal E}_2 |_{\rm RGZ}
=
- \frac{g^2 C_F (3-2 \epsilon) \Gamma(\epsilon-\tfrac{1}{2})}{
2 (8 \pi)^{3/2-\epsilon}
Q^2}
\mu^{2 \epsilon} 
\nonumber \\ & \quad \times 
\bigg[ \frac{M^2-m^2+Q^2}{(m^2+M^2 -Q^2)^{\epsilon-\frac{1}{2}}} 
+ \frac{m^2-M^2+Q^2}{(m^2+M^2 +Q^2)^{\epsilon-\frac{1}{2}}}
\bigg]
\nonumber \\ & \quad 
+O(\alpha_s^2)
\nonumber \\
&= 
\frac{3 g^2 C_F}{16 \sqrt{2} \pi Q^2} 
\Big[ (M^2-m^2+Q^2) \sqrt{m^2+M^2 -Q^2} 
\nonumber \\ & \quad \quad 
+ (m^2-M^2+Q^2) \sqrt{m^2+M^2 +Q^2} \Big] 
\nonumber \\ & \quad
+ O(\alpha_s^2,\epsilon). 
\end{align}
We can see from the first equality that the integral contains a linear power
divergence from the pole at $\epsilon=1/2$. In the second equality, we expanded
in powers of $\epsilon$, which subtracts this power divergence according to
dimensional regularization and we obtain a finite result.

\subsection{Numerical results}
\label{sec:number}

Now we are ready to present numerical results for the moments. 
We first compute ${\cal E}_3$ in the $\overline{\rm MS}$ scheme. 
Because we work at tree level, there is some ambiguity in the choice of the 
strong coupling. Since the parameters of the RGZ theory are obtained from the
lattice data with the lattice coupling around $\beta = 6.0$, it seems
appropriate to compute the coupling from the relation $\beta = 6/g^2$ 
when computing low-energy operator matrix elements. 
We take the numerical values of the parameters $M$, $m$, and $\lambda$ from
Ref.~\cite{Dudal:2010tf} as we have done in the previous section. 
By using Eq.~(\ref{eq:E3integral_renorm}) we obtain
at scale $\Lambda = 1$~GeV the value for ${\cal E}_3$ given by 
\begin{equation}
\label{eq:E3result_1GeV}
{\cal E}_3^{(\Lambda =1\,{\rm GeV})} \big|_{\textrm{RGZ, tree level}} 
= 1.03^{+0.12}_{-0.10}, 
\end{equation}
where the uncertainties come from the parameters in the RGZ theory. 
Note that our result for ${\cal E}_3$ at tree level misses the contribution 
from ${\cal D}(\tau^2)$, which only occurs from order $\alpha_s^2$. 
We can estimate the effect of the inclusion of ${\cal D}(\tau^2)$ at long
distances by using the lattice measurement of ${\cal D}_B (\tau^2)$, 
which includes ${\cal D}(\tau^2)$, and
comparing it with the result from the RGZ theory containing only 
${\cal D}_1(\tau^2)$. 
We neglect the short-distance contribution from ${\cal D}(\tau^2)$ because it
is of higher orders in the strong coupling, and the short-distance behavior 
approximately cancels in the difference between the lattice and the RGZ results
for ${\cal D}_B (\tau^2)$. Hence, we use the long-distance functional forms 
we obtained in Sec.~\ref{sec:QCDcorr} to estimate the long-distance 
contribution from ${\cal D}(\tau^2)$. 
The contribution to ${\cal E}_3$ from ${\cal D}(\tau^2)$ is then 
\begin{align}
\label{eq:E3add_RGZ}
&
- \frac{d-1}{N_c}
\int_0^\infty d \tau \, \tau^3 {\cal D} (\tau^2) 
\nonumber \\ & \quad 
\approx 
- 
\int_0^\infty d \tau \, \tau^3 
\left( A e^{-\tau/\lambda_a} - C e^{-\tau/\lambda_c} \right) 
\nonumber \\ & \quad 
= 0.43^{+0.30}_{-0.38}, 
\end{align}
where $A e^{-\tau/\lambda_a}$ comes from the lattice QCD result for 
${\cal D}_B (\tau^2)$ in Ref.~\cite{Bali:1997aj} and $C e^{-\tau/\lambda_c}$ is
the long-distance behavior of the RGZ result for ${\cal D}_B (\tau^2)$
determined in Sec.~\ref{sec:QCDcorr}. 
Alternatively, we could compute the same quantity by obtaining 
${\cal D}_1 (\tau^2)$ from lattice measurements of 
$D_* (\tau^2) = \tau^2 \frac{\partial}{\partial \tau^2} {\cal D}_1 (\tau^2)$, 
whose long-distance behavior is given in Ref.~\cite{Bali:1997aj} by 
$-B e^{-\tau/\lambda_b}$. 
From this we obtain an expression for ${\cal D}_1 (\tau^2)$ in terms of the 
exponential integral 
${\cal D}_1^{\rm lat} (\tau^2) = 2 B \, E_1(\tau/\lambda_b)$, 
where $E_1 (z) = \int_z^\infty dt\, e^{-t}/t$. 
Note that this functional form behaves at large $\tau$ like 
${\cal D}_1^{\rm lat} (\tau^2) \approx 
2 B \lambda_b e^{-\tau/\lambda_b} /\tau$, 
which is different from the functional form used for ${\cal D}_B (\tau^2)$. 
From these we obtain the contribution to ${\cal E}_3$ from ${\cal D} (\tau^2)$
given by 
\begin{align}
\label{eq:E3add_lat}
& 
- \frac{d-1}{N_c} \int_0^\infty d \tau \, \tau^3 {\cal D} (\tau^2)
\nonumber \\ & \quad \approx 
- \int_0^\infty d \tau \, \tau^3 
\left( A e^{-\tau/\lambda_a} - 2 B \, E_1 (\tau/\lambda_b) \right)
\nonumber \\ & \quad 
= 0.42^{+0.69}_{-0.71}.
\end{align}
Remarkably, this result has a central value that is very close to the result in
Eq.~(\ref{eq:E3add_RGZ}), but with much larger uncertainty. 
We take the result in Eq.~(\ref{eq:E3add_RGZ}) as our estimate for the 
contribution from ${\cal D}(\tau^2)$ to ${\cal E}_3$. 
We combine the results in 
Eqs.~(\ref{eq:E3result_1GeV}) and (\ref{eq:E3add_RGZ}) to obtain 
\begin{equation}
\label{eq:E3result_1GeV_final}
{\cal E}_3^{(\Lambda =1\,{\rm GeV})} 
= 1.46 ^{+0.32}_{-0.39}, 
\end{equation}
where the uncertainties are added in quadrature. Note that the uncertainties
are dominated by the unknown ${\cal D} (\tau^2)$. 
This is our final numerical result for ${\cal E}_3$ at the $\overline{\rm MS}$
scale $\Lambda =1$~GeV. 

In the case of ${\cal E}_1$, we have established that the contribution from
${\cal D}_1(\tau^2)$ vanishes not just in the RGZ theory, but in general, 
as long as power divergences are properly subtracted in
dimensional regularization. Hence, the only contribution to ${\cal E}_1$ comes
from ${\cal D}(\tau^2)$. We can again estimate this contribution by comparing 
our result for ${\cal D}_B (\tau^2)$ in the RGZ theory with the lattice 
measurement. We have 
\begin{align}
\label{eq:E1_RGZ}
{\cal E}_1 &= \frac{d-1}{N_c}
\int_0^\infty d \tau \, \tau {\cal D} (\tau^2) 
\nonumber \\ & 
\approx
\int_0^\infty d \tau \, \tau
\left( A e^{-\tau/\lambda_a} - C e^{-\tau/\lambda_c} \right)
\nonumber \\ & 
= 0.01^{+0.13}_{-0.09}~{\rm GeV}^2. 
\end{align}
If instead we use the lattice measurement for $D_* (\tau^2)$ in
Ref.~\cite{Bali:1997aj} to obtain ${\cal D}_1(\tau^2)$, then we obtain
\begin{align}
\label{eq:E1_lat}
{\cal E}_1|_{\rm lat} &\approx
\int_0^\infty d \tau \, \tau
\left[ A e^{-\tau/\lambda_a} - 2 B \, E_1 (\tau/\lambda_b) \right]
\nonumber \\ & = -0.08^{+0.19}_{-0.21}~{\rm GeV}^2,
\end{align}
which is compatible with Eq.~(\ref{eq:E1_RGZ}), but with a larger
uncertainty. Note that neither results are precise enough to determine the 
sign. We take Eq.~(\ref{eq:E1_RGZ}) as our estimate for ${\cal E}_1$ 
because the uncertainty is smaller. 

Next, ${\cal B}_1$ in the RGZ theory at tree level can be computed from 
Eq.~(\ref{eq:B1result_AAfinite}). We obtain
\begin{equation}
\label{eq:B1RGZ_number}
{\cal B}_1|_{\textrm{RGZ, tree level}} = 1.02 \pm 0.08~{\rm GeV}^2,  
\end{equation}
where the uncertainty comes from the uncertainty in $m^2$. 
Similarly to ${\cal E}_3$ and ${\cal E}_1$, this result is missing the
contribution from ${\cal D}(\tau^2)$. 
By adding this contribution, which is equal to our estimate for ${\cal E}_1$ in 
Eq.~(\ref{eq:E1_lat}), we obtain
\begin{equation}
\label{eq:B1RGZ_numberfinal}
{\cal B}_1 = 1.03^{+0.15}_{-0.12}~{\rm GeV}^2,
\end{equation}
which is our final numerical result for ${\cal B}_1$. 

Finally, we compute $i{\cal E}_2$ in the RGZ theory:
\begin{equation}
\label{eq:E2RGZ_number}
i {\cal E}_2|_{\textrm{RGZ, tree level}} = -0.18 \pm 0.03~{\rm GeV}, 
\end{equation}
where the uncertainties come from the parameters in the RGZ theory. 
Similarly to ${\cal E}_3$,  ${\cal E}_1$, and ${\cal B}_1$, this result is
missing the contribution from ${\cal D} (\tau^2)$. 
We estimate the contribution from ${\cal D} (\tau^2)$ to $i {\cal E}_2$ 
in the same way as we have done for ${\cal E}_3$. We have 
\begin{align}
\label{eq:E2add_RGZ}
& 
\frac{d-1}{N_c}
\int_0^\infty d \tau \, \tau^2 {\cal D} (\tau^2)
\nonumber \\ \quad 
& \approx
-
\int_0^\infty d \tau \, \tau^2
\left( A e^{-\tau/\lambda_a} - C e^{-\tau/\lambda_c} \right)
\nonumber \\ \quad & 
= -0.10^{+0.18}_{-0.14},
\end{align}
If we instead use only the lattice QCD results in Ref.~\cite{Bali:1997aj}
to estimate this contribution, then we obtain $-0.13 \pm 0.30$, which is consistent
with above but with larger uncertainties. We combine
Eqs.~(\ref{eq:E2RGZ_number}) and (\ref{eq:E2add_RGZ}) to obtain our final
numerical result for $i {\cal E}_2$:
\begin{equation}
\label{eq:E2_result}
i {\cal E}_2 = -0.28^{+0.18}_{-0.14}~{\rm GeV}.
\end{equation}

We may compare our numerical results based on the RGZ theory
with estimates based on the lattice QCD analysis in Ref.~\cite{Bali:1997aj}. 
In the case of ${\cal E}_1$, we would obtain the same result as in
Eq.~(\ref{eq:E1_lat}), since the contribution from ${\cal D}_1(\tau^2)$ 
cancels in ${\cal E}_1$ and we obtain the contribution from ${\cal D}(\tau^2)$ 
by using the lattice results in Ref.~\cite{Bali:1997aj}. 
In the case of ${\cal B}_1$, we may use the long-distance
functional form of ${\cal D}_B(\tau^2)$ given in Ref.~\cite{Bali:1997aj} by 
$D_\perp(\tau^2) = A e^{-\tau/\lambda_a}$ to obtain 
\begin{equation}
\label{eq:B1_lat}
{\cal B}_1|_{\rm lat} = 
\int_0^\infty d \tau \, 
\tau A e^{-\tau/\lambda_a}
= 0.35^{+0.13}_{-0.09} \, {\rm GeV}^2. 
\end{equation}
This result is much smaller than what we obtained in
Eq.~(\ref{eq:B1RGZ_numberfinal}) based on the RGZ theory. 
Note that because we used the long-distance functional form used in the lattice
QCD analysis, the quadratic power divergence expected from the perturbative QCD
calculation is completely missing in Eq.~(\ref{eq:B1_lat}). 
In the case of the dimensionless moment ${\cal E}_3$, the perturbative
power-law contribution is both UV and IR divergent, so that it is not possible
to obtain a dimensionally regulated result simply from the long-distance
functional forms used in the lattice QCD analysis. That is, if we compute the
moment ${\cal E}_3$ from the long-distance functional forms obtained in the
lattice QCD analysis, the result will be missing the perturbative
order-$\alpha_s$ scheme-dependent finite pieces as well as the logarithmic
dependence on the renormalization scale. Simply computing the
moment from the long-distance functional form of ${\cal D}_E(\tau^2) =
D_\perp(\tau^2) + D_* (\tau^2)$ from Ref.~\cite{Bali:1997aj} gives 
\begin{align}
\label{eq:E3_lat}
{\cal E}_3|_{\rm lat} &=
- \int_0^\infty d \tau \, \tau^3 
\left( A e^{-\tau/\lambda_a} - B e^{-\tau/\lambda_b} \right)
\nonumber \\ & 
= 1.61^{+1.26}_{-1.29}. 
\end{align}
Although the central value is similar in size to the $\overline{\rm
MS}$-renormalized result in the RGZ theory at $\Lambda=1$~GeV in
Eq.~(\ref{eq:E3result_1GeV_final}), 
it is completely ambiguous which to renormalization scheme and scale this 
result corresponds. Also the uncertainty in this lattice-based estimate 
is too large to be phenomenologically useful. 
Similarly, a lattice estimate of $i{\cal E}_2$ gives 
\begin{align}
\label{eq:E2_lat}
i {\cal E}_2|_{\rm lat} &=
\int_0^\infty d \tau \,\tau^2
\left( A e^{-\tau/\lambda_a} - B e^{-\tau/\lambda_b} \right)
\nonumber \\ & 
= -0.41^{+0.39}_{-0.42}~{\rm GeV},
\end{align}
which is compatible with our result in Eq.~(\ref{eq:E2_result}), but with much
larger uncertainties. 

Now that we have obtained our numerical estimates for the dimensionless moment 
${\cal E}_3$ in the $\overline{\rm MS}$ scheme and the dimension-two moments 
${\cal E}_1$ and ${\cal B}_1$ in dimensional regularization, we can proceed to 
discuss phenomenological applications in the following section.

\section{Quarkonium Decays}
\label{sec:decays}

Based on the results for the moments of two-point field-strength correlators in
the previous section, we now compute the color-octet matrix elements that
appear in heavy quarkonium decay rates and discuss phenomenological
applications. 

We begin with the inclusive decays of $\chi_{QJ}$ states ($Q = c$ or $b$),
which involve the dimensionless moment ${\cal E}_3$. The relation between 
${\cal E}_3$ and the color-octet matrix element is given
by~\cite{Brambilla:2001xy, Brambilla:2002nu}
\begin{equation}
\label{eq:E3relation_bare}
\frac{m_Q^2 \langle \chi_{QJ} | {\cal O}_8 (^3S_1) | \chi_{QJ} \rangle}
{\langle \chi_{QJ} | {\cal O}_1 (^3P_J) | \chi_{QJ} \rangle}
= \frac{2 T_F}{3 (d-1) N_c} {\cal E}_3, 
\end{equation}
where $m_Q$ is the heavy quark pole mass, and we take the definitions of
the NRQCD operators ${\cal O}_8 (^3S_1)$ and ${\cal O}_1 (^3P_J)$ in 
Refs.~\cite{Bodwin:1994jh, Brambilla:2002nu}. 
The subscripts $1$ and $8$ stand for the color state of the $Q \bar Q$
that is created and annihilated by the NRQCD operator,
and the spectroscopic notation $^{2 S+1} L_J$ is used to indicate the 
spin, orbital, and total angular momentum state of the $Q \bar Q$. 
The factor $m_Q^2$ is included on
the left-hand side in order to make the ratio dimensionless. 
This relation has been obtained in Refs.~\cite{Brambilla:2001xy,
Brambilla:2002nu} in three spatial dimensions\footnote{The quantity ${\cal E}$
defined in Ref.~\cite{Brambilla:2001xy} corresponds to $N_c {\cal E}_3$. 
The expression in Eq.~(156) of Ref.~\cite{Brambilla:2002nu} applies only to the
spin-singlet state, and the expressions for spin-triplet states can be obtained
from heavy-quark spin symmetry.}. 
Since both $\langle \chi_{QJ} | {\cal O}_8 (^3S_1) | \chi_{QJ} \rangle$ and 
${\cal E}_3$ contain logarithmic UV divergences, the relation must
be obtained in $d=4-2 \epsilon$ spacetime dimensions in order to facilitate
correct renormalization of both quantities in the $\overline{\rm MS}$ scheme; 
the above result valid for any $d$ can be obtained by
rederiving the relation in arbitrary spacetime dimensions. 
It is easy to see that the denominator factor $d-1$ comes from the tensor
$\delta^{ij}/(d-1)$ which arises from taking an average over spatial directions
of a rank-2 Cartesian tensor in Eq.~(73) of Ref.~\cite{Brambilla:2002nu}. 
Due to the UV poles in the unrenormalized color-octet matrix element and the 
unrenormalized moment ${\cal E}_3$, the
following relation holds between the $\overline{\rm MS}$-renormalized 
color-octet matrix element and the moment:
\begin{align}
\label{eq:E3relation_ren}
\rho_8^{(\Lambda)} & \equiv 
\frac{m_Q^2 \langle \chi_{QJ} | {\cal O}_8 (^3S_1) | \chi_{QJ} \rangle^{(\Lambda)}}
{\langle \chi_{QJ} | {\cal O}_1 (^3P_J) | \chi_{QJ} \rangle}
\nonumber \\ & 
= \frac{2 T_F}{9 N_c} \left( {\cal E}_3^{(\Lambda)} + \frac{g^2
C_F}{\pi^2} + O(\alpha_s^2) \right), 
\end{align}
where the extra finite piece comes from the product of the 
order-$\epsilon$ term in the $d$-dependent coefficient 
$3/(d-1) = 1+\frac{2}{3} \epsilon + O(\epsilon^2)$ and the $1/\epsilon$ pole in
the unrenormalized ${\cal E}_3$. Following Refs.~\cite{Bodwin:2007zf,
CLEO:2008bsq} we refer to this ratio as $\rho_8^{(\Lambda)}$. 
In phenomenological determinations of the quantity ${\cal E}_3$, 
such extra finite terms were not taken into account, and so, when comparing
with the $\overline{\rm MS}$-renormalized moment ${\cal E}_3^{(\Lambda)}$,  
the phenomenologically determined ${\cal E}_3$ must be compared with the 
quantity in the parenthesis of Eq.~(\ref{eq:E3relation_ren}) that contains the
extra finite term. 
By using our result for ${\cal E}_3^{(\Lambda=\textrm{1~GeV})}$ in
Eq.~(\ref{eq:E3result_1GeV_final}), we obtain 
\begin{equation}
\label{eq:E3value}
{\cal E}_3^{(\Lambda=\textrm{1~GeV})} + \frac{g^2 C_F}{\pi^2}
= 1.60 ^{+0.32}_{-0.39}. 
\end{equation}
This agrees within uncertainties with a recent phenomenological determination 
$2.05^{+0.94}_{-0.65}$ in Ref.~\cite{Brambilla:2020xod}, 
but with much smaller uncertainties. 
It is straightforward to compute $\chi_{QJ}$ decay rates from this result. 
The NRQCD factorization formula for $\chi_{QJ}$ decay rates to light hadrons 
at leading order in $v$ is given by~\cite{Bodwin:1994jh} 
\begin{align}
& 
\Gamma(\chi_{QJ} \to {\rm LH} ) = 
\frac{2\, {\rm Im} f_1^{(\Lambda)}(^3P_J)}{m^4_Q} 
\langle \chi_{QJ} | {\cal O}_1 (^3P_J) | \chi_{QJ} \rangle 
\nonumber \\ & \quad
+
\frac{2\, {\rm Im} f_8(^3S_1)}{m^2_Q} 
\langle \chi_{QJ} | {\cal O}_8 (^3S_1) | \chi_{QJ} \rangle^{(\Lambda)} , 
\end{align}
where $2\, {\rm Im} f_1^{(\Lambda)}(^3P_J)$ and $2\, {\rm Im} f_8(^3S_1)$ 
are short-distance
coefficients which begin at order $\alpha_s^2$ and are known to
order-$\alpha_s^3$ accuracy~\cite{Petrelli:1997ge}; exceptionally $2 {\rm Im}
f_1^{(\Lambda)}(^3P_1)$ vanishes at order $\alpha_s^2$ because a spin-1 state
cannot decay into two gluons at tree level. 
The short-distance coefficients for the color-singlet channels contain 
logarithmic dependences on $\Lambda$ that cancel the scale dependence of 
the color-octet matrix element. 
While the color-octet matrix element $\langle \chi_{QJ} | {\cal O}_8 (^3S_1) |
\chi_{QJ} \rangle^{(\Lambda)}$ can be reexpressed in terms of 
${\cal E}_3^{(\Lambda)}$ and the 
color-singlet matrix element $\langle \chi_{QJ} | {\cal O}_1 (^3P_J) |
\chi_{QJ} \rangle$ by using Eq.~(\ref{eq:E3relation_ren}), 
the color-singlet matrix element can be computed in terms of quarkonium
wave functions:
\begin{equation}
\label{eq:chiQsinglet}
\langle \chi_{QJ} | {\cal O}_1 (^3P_J) | \chi_{QJ} \rangle
= \frac{3 N_c}{2 \pi} |R'_{\chi_{QJ}}(0)|^2, 
\end{equation}
where $R_{\chi_{QJ}}(r)$ is the radial wave function of the $\chi_{QJ}$ state,
which can be obtained by solving a Schr\"odinger equation. 
Rather than using potential models to compute $|R'_{\chi_{QJ}}(0)|^2$, which
can vary greatly depending on the choice of the model (see, for example, 
Ref.~\cite{Brambilla:2020xod}), we take the first-principles calculations of
the $P$-wave wave functions in Ref.~\cite{Chung:2021efj}. 
Because we include corrections of relative order $\alpha_s$ in the
short-distance coefficients, we include the one-loop
correction to the wave function which comes from the radiative correction to the
static potential, as has been done in Ref.~\cite{Chung:2021efj}. 
We neglect higher-order corrections that were included in
Ref.~\cite{Chung:2021efj} that are associated with two-loop corrections to the
wave function, because that would require including corrections of relative
order $\alpha_s^2$ to the short-distance coefficients that are currently
unknown. 
We also include a part of the relativistic correction that comes from the 
phase space by replacing an overall factor of $1/(2 m_Q)$ by $1/M_{\chi_{QJ}}$
as have been done in Ref.~\cite{Chung:2021efj}. 
Consistently with Ref.~\cite{Chung:2021efj}, we set $m_c = 1.316$~GeV and 
$m_b = 4.743$~GeV, and compute $\alpha_s$ in the
$\overline{\rm MS}$ scheme by using {\sf RunDec} at four-loop accuracy at scales
$\mu = 2.5$~GeV for $Q=c$ and $\mu =5$~GeV for $Q=b$.
We use the quarkonium masses $M_{\chi_{cJ}} = 3.47$~GeV,
$M_{\chi_{bJ}(1P)} = 9.94$~GeV, $M_{\chi_{bJ}(2P)} = 10.26$~GeV, and
$M_{\chi_{bJ}(3P)} = 10.53$~GeV, which were 
computed in Ref.~\cite{Chung:2021efj} and agree within 0.1~GeV with the Particle Data Group (PDG)
values in Ref.~\cite{Workman:2022ynf}. 
In the short-distance coefficients, we set the number of light quark flavors
$n_f$ to be $n_f = 3$ for $Q=c$ and $n_f = 4$ for $Q =b$, 
and fix $\Lambda = 1$~GeV. 
We obtain the following decay rates for $\chi_{cJ}$:
\begin{subequations}
\label{eq:chicJdecays}
\begin{align}
\Gamma(\chi_{c0} \to {\rm LH}) &= 13.5^{+0.1}_{-0.1} {}^{+6.6}_{-3.0} \pm 4.1~{\rm
MeV} \nonumber \\ & 
= 13.5 ^{+7.8}_{-5.1}~{\rm MeV}, \\
\Gamma(\chi_{c1} \to {\rm LH}) &= 0.41 {}^{+0.10}_{-0.12} 
{}^{+0.10}_{-0.07} \pm 0.12~{\rm MeV}
\nonumber \\ & 
= 0.41^{+0.19}_{-0.19}~{\rm MeV}, \\
\Gamma(\chi_{c2} \to {\rm LH}) &= 2.15 {}^{+0.10}_{-0.12} 
{}^{+0.00}_{-0.19} \pm 0.65~{\rm MeV} 
\nonumber \\ & 
= 2.15^{+0.65}_{-0.68}~{\rm MeV}, 
\end{align}
\end{subequations}
where the first uncertainties come from the uncertainty in ${\cal E}_3$, 
the second uncertainties come from varying $\mu$ between 1.5 and 4~GeV,
and the third uncertainties come from uncalculated corrections of order $v^2$,
which we estimate by $0.3$ times the central values, reflecting that 
typically $v^2 \approx 0.3$ for charmonia. 
In the last equalities we add the uncertainties in quadrature. 
These results may be compared directly with experiment; however, the total
widths of $\chi_{cJ}$ include sizable contribution from radiative decays into
$J/\psi + \gamma$, especially for $J=1$ and $2$, which will not be captured by
the calculation of $\Gamma(\chi_{cJ} \to {\rm LH})$. 
After subtracting these radiative decay rates by using the measured branching
fractions, we obtain from Ref.~\cite{Workman:2022ynf} the experimental results  
$\Gamma(\chi_{c0} \to {\rm LH})|_{\rm PDG} = 10.6 \pm 0.6$~MeV, 
$\Gamma(\chi_{c1} \to {\rm LH})|_{\rm PDG} = 0.55 \pm 0.04$~MeV, 
and 
$\Gamma(\chi_{c2} \to {\rm LH})|_{\rm PDG} = 1.60 \pm 0.09$~MeV. 
These are in agreements with our theoretical results in 
Eqs.~(\ref{eq:chicJdecays}), although the experimental values have much smaller
uncertainties. 
When compared with the calculation in Ref.~\cite{Brambilla:2020xod} based on
the phenomenological determination of ${\cal E}_3$, the precision has improved 
for the decay rate of $\chi_{c1}$, while the uncertainty in the decay rate of 
$\chi_{c0}$ is increased. While the improvement in precision has mostly to do
with the improved determination of ${\cal E}_3$ and the calculation of the
wave functions from first principles, the increased uncertainty mainly comes
from the fact that we included an uncertainty from variation of the QCD
renormalization scale, while in Ref.~\cite{Brambilla:2020xod} the
uncertainties from uncalculated corrections of higher orders in $\alpha_s$ were
estimated to be $\alpha_s^2$ times the central value. 
The size of the uncertainty in the decay rate of $\chi_{c2}$ is comparable to
the result in Ref.~\cite{Brambilla:2020xod}. 

Similarly, we obtain the following decay rates for $\chi_{bJ}(1P)$:
\begin{subequations}
\label{eq:chibJ1Pdecays}
\begin{align}
\Gamma(\chi_{b0}(1P) \to {\rm LH}) &= 
0.762 ^{+0.007}_{-0.009} {}^{+0.320}_{-0.066} \pm 0.076 ~{\rm MeV}
\nonumber \\ & 
= 0.762^{+0.329}_{-0.102} ~{\rm MeV},
\\
\Gamma(\chi_{b1}(1P) \to {\rm LH}) &= 
0.052^{+0.007}_{-0.009} {}^{+0.035}_{-0.006} \pm 0.005 ~{\rm MeV}
\nonumber \\ & 
= 0.052^{+0.036}_{-0.012} ~{\rm MeV},
\\
\Gamma(\chi_{b2}(1P) \to {\rm LH}) &= 
0.139^{+0.007}_{-0.009} {}^{+0.004}_{-0.084} \pm 0.014 ~{\rm MeV}
\nonumber \\ & 
= 0.139^{+0.016}_{-0.086} ~{\rm MeV}, 
\end{align}
\end{subequations}
where the first uncertainties come from the uncertainty in ${\cal E}_3$,
the second uncertainties come from varying $\mu$ between 2~GeV and 8~GeV,
and the third uncertainties come from uncalculated corrections of order $v^2$,
which we estimate by $0.1$ times the central values, reflecting that
typically $v^2 \approx 0.1$ for bottomonia.
In the last equalities we add the uncertainties in quadrature.
Unfortunately, experimental results for the total widths of $\chi_{bJ}$ states 
have not been made available yet. 
There are, however, determinations based on the theoretical calculations of the
radiative decay rates into $\Upsilon+\gamma$ and the measured branching
fractions of the same process. 
In Ref.~\cite{Segovia:2018qzb}, the radiative decay rates were computed in 
potential NRQCD at weak coupling, from which the total decay rates were
computed by using the measured branching fractions. 
After subtracting the radiative decay rates we obtain from
Ref.~\cite{Segovia:2018qzb} the values 
$\Gamma(\chi_{b0}(1P) \to {\rm LH})|_{\textrm{Ref.~\cite{Segovia:2018qzb}}} = 
1.4 \pm 0.2$~MeV, 
$\Gamma(\chi_{b1}(1P) \to {\rm LH})|_{\textrm{Ref.~\cite{Segovia:2018qzb}}} = 
0.069 \pm 0.009$~MeV, 
and 
$\Gamma(\chi_{b2}(1P) \to {\rm LH})|_{\textrm{Ref.~\cite{Segovia:2018qzb}}} = 
0.195 \pm 0.021$~MeV. 
While the decay width for the $\chi_{b1}$ state is in agreement with our
results, the results for $\chi_{b0}$ and $\chi_{b2}$ are much larger than what
we obtain in Eqs.~(\ref{eq:chibJ1Pdecays}). 
In Ref.~\cite{Zhang:2023mky}, the authors took the results for the radiative 
decay rates 
computed in a nonrelativistic model based on the Cornell potential available 
in Table~4.16 in Ref.~\cite{QuarkoniumWorkingGroup:2004kpm} and used the
measured results for the branching fractions in Ref.~\cite{Workman:2022ynf} 
to compute the total decay widths. We again subtract the radiative decay rates
from the results listed in Ref.~\cite{Zhang:2023mky} to obtain 
$\Gamma(\chi_{b0}(1P) \to {\rm LH})|_{\textrm{Ref.~\cite{Zhang:2023mky}}} =
1.121^{+0.185}_{-0.140} $~MeV,
$\Gamma(\chi_{b1}(1P) \to {\rm LH})|_{\textrm{Ref.~\cite{Zhang:2023mky}}} =
0.051^{+0.005}_{-0.004}$~MeV,
and
$\Gamma(\chi_{b2}(1P) \to {\rm LH})|_{\textrm{Ref.~\cite{Zhang:2023mky}}} =
0.144^{+0.010}_{-0.009}$~MeV.
These are in good agreement with our results in Eq.~(\ref{eq:chibJ1Pdecays}). 
Note that, however, the uncertainties in the results in
Ref.~\cite{Zhang:2023mky} may be underestimated because they 
reflect the ones in the measured branching fractions only, 
as no uncertainty is given in the model calculation of the
radiative decay rates in Ref.~\cite{QuarkoniumWorkingGroup:2004kpm}. 

We also compute decay rates for $\chi_{bJ}(2P)$ and $\chi_{bJ}(3P)$ states:
\begin{subequations}
\label{eq:chibJ2Pdecays}
\begin{align}
\Gamma(\chi_{b0}(2P) \to {\rm LH}) &=
1.03 ^{+0.01}_{-0.01} {}^{+0.43}_{-0.09} \pm 0.10 ~{\rm MeV}
\nonumber \\ & 
= 1.03^{+0.44}_{-0.14} ~{\rm MeV},
\\
\Gamma(\chi_{b1}(2P) \to {\rm LH}) &=
0.070^{+0.009}_{-0.011} {}^{+0.047}_{-0.008} \pm 0.007 ~{\rm MeV}
\nonumber \\ & 
= 0.070^{+0.049}_{-0.016} ~{\rm MeV},
\\
\Gamma(\chi_{b2}(2P) \to {\rm LH}) &=
0.19^{+0.01}_{-0.01} {}^{+0.01}_{-0.11} \pm 0.019 ~{\rm MeV}
\nonumber \\ & 
= 0.19^{+0.02}_{-0.12} ~{\rm MeV},
\end{align}
\end{subequations}
\begin{subequations}
\label{eq:chibJ3Pdecays}
\begin{align}
\Gamma(\chi_{b0}(3P) \to {\rm LH}) &=
1.19 ^{+0.01}_{-0.01} {}^{+0.50}_{-0.10} \pm 0.12 ~{\rm MeV}
\nonumber \\ & 
= 1.19^{+0.51}_{-0.16} ~{\rm MeV},
\\
\Gamma(\chi_{b1}(3P) \to {\rm LH}) &=
0.081^{+0.011}_{-0.013} {}^{+0.055}_{-0.009} \pm 0.008 ~{\rm MeV}
\nonumber \\ & 
= 0.081^{+0.057}_{-0.018} ~{\rm MeV},
\\
\Gamma(\chi_{b2}(3P) \to {\rm LH}) &=
0.22^{+0.01}_{-0.01} {}^{+0.01}_{-0.13} \pm 0.02 ~{\rm MeV}
\nonumber \\ & 
= 0.22^{+0.03}_{-0.13} ~{\rm MeV}.
\end{align}
\end{subequations}
Compared to the results in Ref.~\cite{Brambilla:2020xod} based on the
phenomenological determination of ${\cal E}_3$, 
the decay rates for $J=0$ and 2 are in agreement within uncertainties. 
Although the results for $J=1$ also agree within uncertainties with 
Ref.~\cite{Brambilla:2020xod}, we obtain smaller central values for the decay
rates. The uncertainties in the $\chi_{bJ}$ decay rates are comparable in size
with the results in Ref.~\cite{Brambilla:2020xod}; however, the uncertainties
would have been much smaller if we estimated the uncertainties from
uncalculated corrections of higher orders in $\alpha_s$ in the same way as 
Ref.~\cite{Brambilla:2020xod}, instead of varying the QCD renormalization
scale.

In Refs.~\cite{Bodwin:2007zf, CLEO:2008bsq}, the dimensionless ratio
$\rho_8^{(\Lambda)}$ defined in
Eq.~(\ref{eq:E3relation_ren}) for the bottomonium state at the scale 
$\Lambda = 4.6$~GeV was investigated by computing the decays $\chi_{bJ} \to c+X$
and comparing them with measurements of branching fractions ${\rm Br}
(\chi_{bJ} \to D_0 + X)$. If we take into account the evolution of the
color-octet matrix element, then our results lead to 
$\rho_8^{(\Lambda = 4.6{\rm ~GeV})} = 0.17 \pm 0.01$. 
This is compatible with the result 
$0.160^{+0.071}_{-0.047}$ obtained in Ref.~\cite{CLEO:2008bsq} from the
measured branching fractions for the $1P$ states. In contrast, 
Ref.~\cite{CLEO:2008bsq} obtained a smaller result $0.074^{+0.010}_{-0.008}$
from measurements involving $2P$ states, while we expect
Eq.~(\ref{eq:E3relation_ren}) to be approximately independent of the radial
excitation; however, we note that the quality of the fit for the $2P$ states 
in Ref.~\cite{CLEO:2008bsq} is much worse compared to the $1P$ states, 
resulting in a value of $\chi^2$ that is more than an order of magnitude larger
than that of the $1P$ case. 

\begin{figure}
\includegraphics[width=0.49\textwidth]{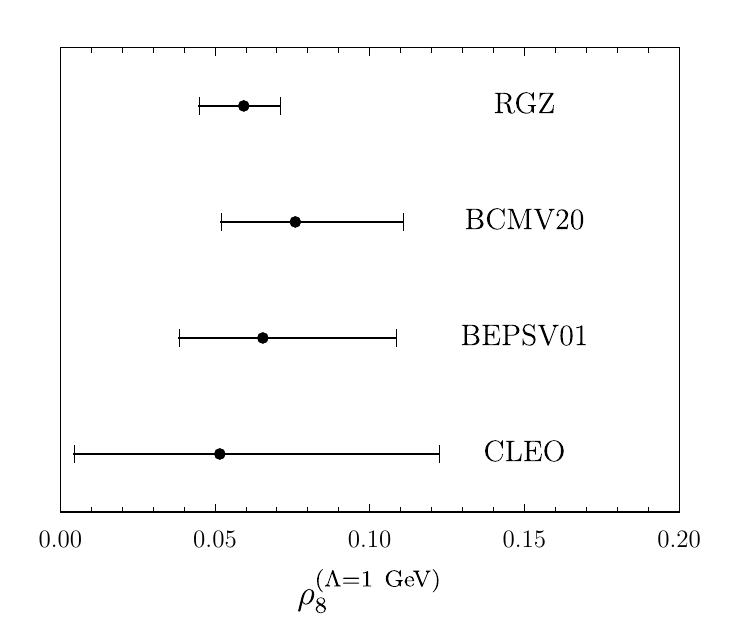}
\caption{\label{figrho}
Results for the $\chi_{QJ}$ matrix element ratio $\rho_8^{(\Lambda)}$ defined in 
Eq.~(\ref{eq:E3relation_ren}) at $\Lambda = 1$~GeV in this work (RGZ), 
compared to determinations based on $\chi_{cJ}$ decay rates in
Ref.~\cite{Brambilla:2001xy} (BEPSV01) and Ref.~\cite{Brambilla:2020xod}
(BCMV20), and the determination based on measurements of ${\rm Br}
(\chi_{bJ}(1P) \to D_0 + X)$ in Ref.~\cite{CLEO:2008bsq} (CLEO). 
}
\end{figure}
In Fig.~\ref{figrho} we compare our result for the $\chi_{QJ}$ matrix element
ratio $\rho_8^{(\Lambda)}$ defined in
Eq.~(\ref{eq:E3relation_ren}) at $\Lambda = 1$~GeV with the phenomenological
determinations in Ref.~\cite{Brambilla:2001xy} (BEPSV01) and 
Ref.~\cite{Brambilla:2020xod} (BCMV20) based on
$\chi_{cJ}$ decay rates, and the determination in Ref.~\cite{CLEO:2008bsq}
(CLEO) from measurements of ${\rm Br} (\chi_{bJ}(1P) \to D_0 + X)$. 
Note that because the result in Ref.~\cite{CLEO:2008bsq} has been obtained at 
the scale of the bottom quark mass, we solved the evolution equation for
${\cal E}_3$ in Eq.~(\ref{eq:E3RG}) to obtain the result at $\Lambda = 1$~GeV.
We can see that from Fig.~\ref{figrho}, our result based on the RGZ theory has 
much smaller uncertainties compared to the phenomenological determinations
based on measured decay rates, although the results are consistent within
uncertainties.

We now list our numerical results for the color-octet matrix elements for
$\chi_{cJ}$ and $\chi_{bJ}$ states obtained from the RGZ result for the moment 
${\cal E}_3$ and the calculation of the $P$-wave quarkonium wave functions in
Ref.~\cite{Chung:2021efj} at one-loop level: 
\begin{subequations}
\label{eq:chi8matrixelements}
\begin{align}
\langle {\cal O}_8 (^3S_1) \rangle_{\chi_{cJ}}^{(\Lambda=1~{\rm GeV})} &=
(2.66^{+0.53}_{-0.65}) \times 10^{-3} {\rm ~GeV}^3,
\\
\langle {\cal O}_8 (^3S_1) \rangle_{\chi_{bJ}(1P)}^{(\Lambda=1~{\rm GeV})} &= 
(3.45^{+0.69}_{-0.84}) \times 10^{-3} {\rm ~GeV}^3, 
\\
\langle {\cal O}_8 (^3S_1) \rangle_{\chi_{bJ}(2P)}^{(\Lambda=1~{\rm GeV})} &= 
(4.79^{+0.96}_{-1.17}) \times 10^{-3} {\rm ~GeV}^3, 
\\
\langle {\cal O}_8 (^3S_1) \rangle_{\chi_{bJ}(3P)}^{(\Lambda=1~{\rm GeV})} &=
(5.69^{+1.14}_{-1.39}) \times 10^{-3} {\rm ~GeV}^3. 
\end{align}
\end{subequations}
Here we use the shorthand $\langle {\cal O}_8 (^3S_1)
\rangle_{\chi_{QJ}}^{(\Lambda)}
\equiv \langle \chi_{QJ} | {\cal O}_8 (^3S_1) | \chi_{QJ} \rangle^{(\Lambda)}
$. 
These results are much more precise compared to previous phenomenological 
determinations in Ref.~\cite{Brambilla:2020xod}, with uncertainties reduced 
by a factor of about $1/3$. Note that all of the color-octet matrix elements in 
Eq.~(\ref{eq:chi8matrixelements}) are computed in the $\overline{\rm MS}$
scheme at scale $\Lambda =1$~GeV, and any direct comparison with other results
must be done at the same scheme and scale.

Next we look into the decays of $S$-wave states into light hadrons. 
In the case of $\eta_Q$, where $Q = c$ or $b$, the following relations
hold for the color-octet matrix elements and the dimension-two
moments~\cite{Brambilla:2002nu}:
\begin{subequations}
\label{eq:etaQoctet}
\begin{align}
\label{eq:3S1relation}
\frac{\langle \eta_Q | {\cal O}_8 (^3S_1) | \eta_Q \rangle}
{\langle \eta_Q | {\cal O}_1 (^1S_0) | \eta_Q \rangle}
&= - \frac{c_F^2 {\cal B}_1}{2 m_Q^2 N_c}, 
\\
\label{eq:1P1relation}
\frac{\langle \eta_Q | {\cal O}_8 (^1P_1) | \eta_Q \rangle/m_Q^2}
{\langle \eta_Q | {\cal O}_1 (^1S_0) | \eta_Q \rangle}
&= -\frac{{\cal E}_1}{2 m_Q^2 N_c}, 
\end{align}
\end{subequations}
where $c_F  =1 +O(\alpha_s)$ is the short-distance coefficient in the NRQCD
Lagrangian associated with the spin-flip interaction~\cite{Eichten:1990vp,
Czarnecki:1997dz, Grozin:2007fh}; even though $c_F$ is known to three loops,
we only need the result at tree level consistently with our tree-level
evaluation of ${\cal B}_1$. We take the definitions of the operators 
${\cal O}_8 (^3S_1)$, ${\cal O}_1 (^1S_0)$, and ${\cal O}_8 (^1P_1)$ 
in Refs.~\cite{Bodwin:1994jh, Bodwin:2002cfe, Brambilla:2002nu}. 
We have included a factor of $1/m_Q^2$ on the left-hand side
of Eq.~(\ref{eq:1P1relation}) in order to make the ratio dimensionless. 
Similarly to the $\chi_{QJ}$ case, these ratios do not depend on the radial
excitation of the $\eta_Q$ state, and the only dependence on the heavy quark
flavor comes from the heavy quark mass $m_Q$. 
Note that Eq.~(\ref{eq:1P1relation}) is taken from 
Eq.~(155) in Ref.~\cite{Brambilla:2002nu}, where one needs to set $J=1$ 
for the relation to hold for the spin-singlet case. 
There is an additional color-octet matrix element 
$\langle \eta_Q | {\cal O}_8 (^1S_0) | \eta_Q \rangle$ that enters the decay
rate of spin-singlet $S$-wave quarkonium into light hadrons; we do not consider
this matrix element because it is proportional to a moment of the four-point
field-strength correlation function [see Eqs.~(80) and (153) of
Ref.~\cite{Brambilla:2002nu}], which is outside the scope of this work. 
Before we consider the decay rate of $\eta_Q$ let us compare
Eqs.~(\ref{eq:etaQoctet}) with the estimates of the sizes of the color-octet 
matrix elements given in Ref.~\cite{Bodwin:2002cfe}. 
In Ref.~\cite{Bodwin:2002cfe}, the ratio in Eq.~(\ref{eq:3S1relation}) 
has been estimated to be the order of $v^3/(2 N_c)$, which corresponds to about
0.03 for charmonium, and 0.005 for bottomonium. By using our result for ${\cal
B}_1$ in Eq.~(\ref{eq:B1RGZ_numberfinal}), we obtain $-0.1$ and $-0.008$ for
the right-hand side of Eq.~(\ref{eq:3S1relation}) for charmonium and
bottomonium, respectively. These results are larger in size than the estimates
given in Ref.~\cite{Bodwin:2002cfe}, and the signs are negative due to the
positivity of ${\cal B}_1$. In the case of Eq.~(\ref{eq:1P1relation}), 
this ratio was estimated in Ref.~\cite{Bodwin:2002cfe} to be about 
$v^4/(2 N_c)$, which gives $0.015$ and $0.0017$ for charmonium and bottomonium,
respectively. If we use our result for ${\cal E}_1$ in Eq.~(\ref{eq:E1_RGZ}), 
then the ratio ranges between $-0.014$ and $0.008$ for charmonium, 
and between $-0.0010$ and $0.0006$ for bottomonium, which are smaller than 
the estimates in Ref.~\cite{Bodwin:2002cfe}. 

The sizes of the contributions from the color-octet matrix elements in 
Eqs.~(\ref{eq:etaQoctet}) to the decay rates $\Gamma(\eta_Q \to {\rm LH})$ 
can be computed by multiplying the corresponding short-distance coefficients.
Numerical sizes of the tree-level short-distance coefficients compared to the
one at leading order in $v$ are listed in Ref.~\cite{Bodwin:2002cfe}. 
Compared to the decay rate at leading order $v$, which only involves the
color-singlet matrix element $\langle \eta_Q | {\cal O}_1 (^1S_0) | \eta_Q
\rangle$, the relative contribution from the color-octet matrix element 
$\langle \eta_Q | {\cal O}_8 (^3S_1) | \eta_Q \rangle$
is given by Eq.~(\ref{eq:3S1relation}) times $0.75 n_f$. If we put $n_f = 3$
for charmonium and $n_f = 4$ for bottomonium, then 
the correction to the decay rate from ${\cal B}_1$ is about $-22 \pm 3$\% for
$\eta_c$, and about $-2.3 \pm 0.3$\% for $\eta_b$. 
In the case of the matrix element $\langle \eta_Q | {\cal O}_8 (^1P_1) | \eta_Q
\rangle$, the contribution from ${\cal E}_1$ relative to the leading-order
decay rate is given by $1.13$ times Eq.~(\ref{eq:1P1relation}). 
Because of the small size of ${\cal E}_1$, the contribution from ${\cal E}_1$
to the decay rate is at most of the order of 1\% for $\eta_c$, and is at most
of the order of $0.1$\% for $\eta_b$. 
The order-$v^4$ correction from the matrix element
$\langle \eta_Q | {\cal O}_8 (^1S_0) | \eta_Q \rangle$
involving the four-point correlation function would be of the order of 3\% for
charmonium and 0.3\% for bottomonium according to the estimate given in
Ref.~\cite{Bodwin:2002cfe}. Hence, the only significant color-octet
contribution to the $\eta_Q$ decay rate comes from ${\cal B}_1$. 

In order to properly quantify the color-octet contribution to the $\eta_Q$
decay rate, we need to include radiative corrections to the short-distance
coefficients, which are known to converge slowly~\cite{Bodwin:2001pt}. 
In Ref.~\cite{Brambilla:2018tyu}, the authors computed the ratio 
$R_{\eta_Q} = \Gamma(\eta_Q \to {\rm LH})/\Gamma(\eta_Q \to \gamma \gamma)$ 
by resumming a class of radiative corrections in the large-$n_f$ limit, 
based on an earlier resummation calculation in Ref.~\cite{Bodwin:2001pt} and
a fixed-order calculation at two-loop accuracy in Ref.~\cite{Feng:2017hlu}.
It was found in this work that the lack of knowledge of the color-octet matrix
element $\langle \eta_Q | {\cal O}_8 (^3S_1) | \eta_Q \rangle$
is a significant source of uncertainty. 
Since we can determine this matrix element from our result for 
${\cal B}_1$, we can remove this uncertainty.
Since the calculation in Ref.~\cite{Brambilla:2018tyu} was done by cutting off
the quadratic power divergence in the color-octet matrix element with a hard
cutoff, first we need to convert our result for ${\cal B}_1$ from dimensional
regularization to cutoff regularization. 
The power divergence in ${\cal B}_1$ is simply given by a scaleless 
power-divergent integral which can be read off from Eq.~(\ref{eq:B1integral})
by setting all dimensionful parameters to zero: 
\begin{equation}
\label{eq:B1conversion}
{\cal B}_1^{\rm cutoff} - 
{\cal B}_1^{\rm DR}
= \frac{g^2 C_F}{2 \pi^2} \int_0^\Lambda dk \, k, 
\end{equation}
where $\Lambda$ is the hard cutoff on the spatial loop momentum, consistently
with what was done in Ref.~\cite{Brambilla:2018tyu}. At $\Lambda = 1$~GeV, 
which was used in Ref.~\cite{Brambilla:2018tyu} for charmonium, 
Eq.~(\ref{eq:B1conversion}) amounts to $0.15$~GeV$^2$ if we set $\alpha_s(m_c)
= 0.35$. 
At $\Lambda = 2$~GeV, which was used in Ref.~\cite{Brambilla:2018tyu} for
bottomonium, the cutoff-regulated ${\cal B}_1$ is larger than the dimensionally
regulated one by $0.37$~GeV$^2$ if we use $\alpha_s (m_b) = 0.22$. 
We can now improve the result in Ref.~\cite{Brambilla:2018tyu}
by including the color-octet contribution we compute with our result for 
${\cal B}_1$. 
We first compute $R_{\eta_c}$ for the charmonium states. 
We obtain, from Eq.~(66) of Ref.~\cite{Brambilla:2018tyu}, 
\begin{align}
\label{eq:RetacNNA}
R_{\eta_c} {\rm (NNA)} &= (5.83^{+1.29}_{-0.53}{}^{+0.20}_{-0.16}) \times 10^3
\nonumber \\ &
= (5.83^{+1.31}_{-0.55}) \times 10^3, 
\end{align}
where we shifted the central value by the correction from the
color-octet matrix element. The first uncertainty comes from the variation of
the QCD renormalization scale between 1 and 2~GeV, and the second uncertainty
comes from ${\cal B}_1$. Because we have now included the effect of the
color-octet matrix element in the correct scheme, we have removed the 
uncertainties arising from varying the cutoff $\Lambda$ and from the unknown 
color-octet matrix element. 
Here, NNA refers to the na\"ive non-Abelianization scheme used for the
resummation of radiative corrections in the large-$n_f$
limit~\cite{Beneke:1994qe}. 
The size of the correction from the color-octet matrix element can be easily
obtained by replacing the estimate 
${\langle \eta_Q | {\cal O}_8 (^3S_1) | \eta_Q \rangle}/
{\langle \eta_Q | {\cal O}_1 (^1S_0) | \eta_Q \rangle}
= v^3 C_F/(\pi N_c)$ with $v^2 = 0.3$ used in Ref.~\cite{Brambilla:2018tyu} 
with the pNRQCD result Eq.~(\ref{eq:3S1relation}) 
and by using our result for ${\cal B}_1$. Here, we take $m_c = 1.5$~GeV
consistently with Ref.~\cite{Brambilla:2018tyu}. 
The uncertainties are slightly reduced compared to the result in
Ref.~\cite{Brambilla:2018tyu}, because the uncertainty in the NNA result is
dominated by the variation of the QCD renormalization scale. 
Similarly, from Eq.~(67) of Ref.~\cite{Brambilla:2018tyu} we obtain 
\begin{align}
R_{\eta_c} {\rm (BFG)} &= (5.18^{+0.06}_{-0.18}{}^{+0.23}_{-0.18}) \times 10^3
\nonumber\\ &
= (5.18^{+0.23}_{-0.26}) \times 10^3,
\end{align}
which again we obtain by shifting the central value by the correction from the
color-octet matrix element. Here, BFG refers to the 
background-field gauge method that was used for the
resummation of radiative corrections in the large-$n_f$
limit~\cite{DeWitt:1967ub}.
The sources of uncertainties are same as in Eq.~(\ref{eq:RetacNNA}). 
The uncertainty in the BFG result is greatly reduced compared to the result in
Ref.~\cite{Brambilla:2018tyu}, because the uncertainty was dominated by the
unknown color-octet matrix element. 
Note that, compared to the results in Ref.~\cite{Brambilla:2018tyu}, the
differences between the results in the NNA and BFG methods have reduced by 
inclusion of the correction from the color-octet matrix element, 
and the two results are still in agreement despite the significant reduction of
uncertainties. 
We note that because the order-$v^4$ corrections from color-singlet matrix
elements also cancel in the ratio $R_{\eta_Q}$~\cite{Bodwin:2002cfe}, 
the uncertainties from 
uncalculated order-$v^4$ corrections come from the color-octet matrix elements 
$\langle \eta_Q | {\cal O}_8 (^1P_1) | \eta_Q \rangle$
and $\langle \eta_Q | {\cal O}_8 (^1S_0) | \eta_Q \rangle$. 
These corrections are estimated to be about a few percent, which are well
within the uncertainties in our results. 

In order to compare with experiment, 
we obtain the branching fraction ${\rm Br} (\eta_c \to \gamma \gamma)$ by
taking the inverse of $R_{\eta_c}$. We have 
\begin{subequations}
\label{eq:Bretac}
\begin{align}
{\rm Br} (\eta_c \to \gamma \gamma)|_{\rm NNA} &= (1.71^{+0.18}_{-0.31}) \times
10^{-4}, \\
{\rm Br} (\eta_c \to \gamma \gamma)|_{\rm BFG} &= (1.93^{+0.10}_{-0.08}) \times
10^{-4}. 
\end{align}
\end{subequations}
The NNA result for the branching fraction agrees within uncertainties with the 
PDG value ${\rm Br} (\eta_c \to \gamma \gamma)|_{\rm PDG} = (1.68 \pm 0.12) 
\times 10^{-4}$~\cite{Workman:2022ynf}, but the BFG result is slightly larger. 
Interestingly, a recent lattice QCD determination of the two-photon rate in
Ref.~\cite{Colquhoun:2023zbc}, when divided by the measured total width of
$\eta_c$, leads to the value ${\rm Br} (\eta_c \to \gamma \gamma) =
(2.121 \pm 0.050) \times 10^{-4}$, which is in tension with the PDG value
and is slightly larger than the improved BFG result. 
We note that the results in Eq.~(\ref{eq:Bretac}) are independent of the 
radial excitation and apply
to the $\eta_c(2S)$ state as well. Indeed, the results are in agreement with
the PDG value ${\rm Br} (\eta_c(2S) \to \gamma \gamma)|_{\rm PDG} = (1.6 \pm
1.0) \times 10^{-4}$, although the experimental uncertainties are much larger
than the $1S$ case. 

We can repeat the same analysis for the ratio 
$R_{\eta_b} = \Gamma(\eta_b \to {\rm LH})/\Gamma(\eta_b \to \gamma \gamma)$ for
the bottomonium states. 
Although the relative size of the correction from the color-octet matrix
element is expected to be much smaller than the charmonium case, 
inclusion of the color-octet matrix elements will still be able to reduce the
uncertainties in the results. 
Similarly to the charmonium case, we include the correction from ${\cal B}_1$
by replacing the estimate 
${\langle \eta_Q | {\cal O}_8 (^3S_1) | \eta_Q \rangle}/
{\langle \eta_Q | {\cal O}_1 (^1S_0) | \eta_Q \rangle}
= v^3 C_F/(\pi N_c)$ with $v^2=0.1$ used in Ref.~\cite{Brambilla:2018tyu} with
the pNRQCD result Eq.~(\ref{eq:3S1relation})
and by using our result for ${\cal B}_1$.
We obtain from Eqs.~(68) and (69) of Ref.~\cite{Brambilla:2018tyu} the
following results 
\begin{subequations}
\label{eq:Retab}
\begin{align}
R_{\eta_b} {\rm (NNA)} &= (2.47^{+0.02}_{-0.05}{}^{+0.02}_{-0.01}) \times 10^4
\nonumber \\ & 
= (2.47^{+0.03}_{-0.05}) \times 10^4, \\
R_{\eta_b} {\rm (BFG)} &= (2.58^{+0.00}_{-0.05}{}^{+0.02}_{-0.01}) \times 10^4 
\nonumber \\ & 
= (2.58^{+0.02}_{-0.05}) \times 10^4, 
\end{align}
\end{subequations}
where the uncertainties are as in Eq.~(\ref{eq:RetacNNA}). 
The uncertainties are now reduced by half compared to
the results in Ref.~\cite{Brambilla:2018tyu}. 
The difference between the NNA and BFG results are now slightly larger after
inclusion of the color-octet matrix element, and the two results are now in
slight tension. By taking the inverse of $R_{\eta_b}$, we obtain the branching
fraction 
\begin{subequations}
\begin{align}
{\rm Br} (\eta_b \to \gamma \gamma)|_{\rm NNA} &= (4.05^{+0.09}_{-0.04}) \times
10^{-5}, \\
{\rm Br} (\eta_b \to \gamma \gamma)|_{\rm BFG} &= (3.88^{+0.08}_{-0.03}) \times
10^{-5}.
\end{align}
\end{subequations}
Since the two-photon rates of $\eta_b$ states have not been measured, it is not
possible to compare these results directly with experiment. 
Instead, we compute the total decay widths of the $\eta_b$ states by using the
two-photon rates computed in Ref.~\cite{Chung:2020zqc}. We obtain
\begin{subequations}
\begin{align}
\label{eq:etabtotalrates}
\Gamma_{\eta_b(1S)} &= 10.9^{+4.2}_{-1.7}~{\rm MeV}, 
\\
\Gamma_{\eta_b(2S)} &= 4.9\pm 0.6~{\rm MeV}, 
\\
\Gamma_{\eta_b(3S)} &= 3.6\pm 0.4~{\rm MeV}, 
\end{align}
\end{subequations}
where the central values are obtained from the average of NNA and BFG results,
and the uncertainties come from the improved results for $R_{\eta_b}$ and 
the two-photon rates in Ref.~\cite{Chung:2020zqc}. In all cases, the
uncertainties are dominated by the uncertainties in the predictions for the
two-photon rates. The result for $\Gamma_{\eta_b(1S)}$ is in agreement with the
PDG value $\Gamma_{\eta_b(1S)} = 10^{+5}_{-4}$~MeV~\cite{Workman:2022ynf}. 

\begin{figure*}
\includegraphics[width=0.99\textwidth]{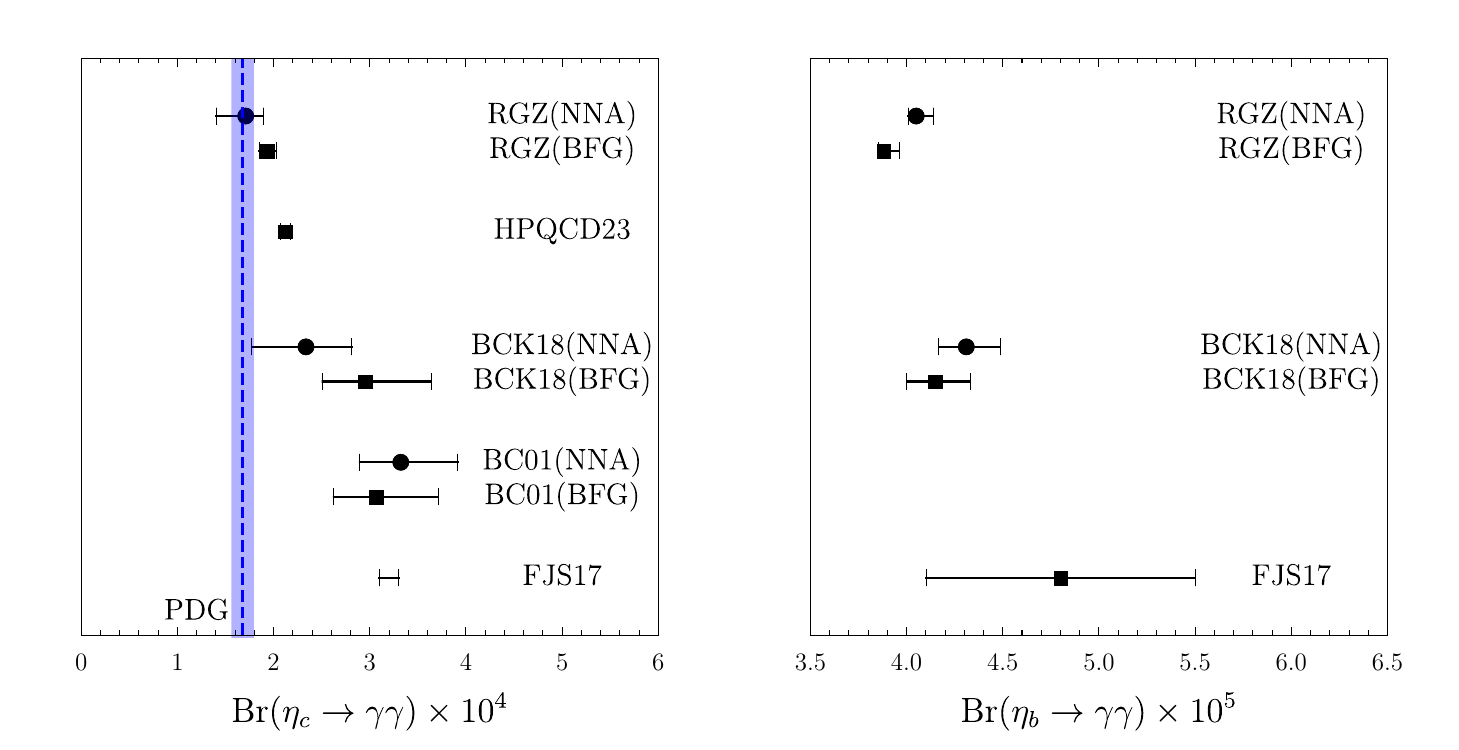}
\caption{\label{figetaQ}
Results for two-photon branching fractions of $\eta_c$ (left) and $\eta_b$
(right) in this work (RGZ) compared to a lattice QCD study in
Ref.~\cite{Colquhoun:2023zbc} (HPQCD23), large-$n_f$ resummation calculations
in Ref.~\cite{Brambilla:2018tyu} (BCK18) and Ref.~\cite{Bodwin:2001pt} (BC01), 
a fixed-order calculation at two-loop accuracy in Ref.~\cite{Feng:2017hlu}
(FJS17). Results for the $\eta_b$ case are not available in
HPQCD23~\cite{Colquhoun:2023zbc} and BC01~\cite{Bodwin:2001pt}.
The PDG value for the $\eta_c$ case from Ref.~\cite{Workman:2022ynf} is shown 
as a blue band. 
}
\end{figure*}

In Fig.~\ref{figetaQ} we compare our results for ${\rm Br} (\eta_c \to \gamma
\gamma)$ and ${\rm Br} (\eta_b \to \gamma \gamma)$ with the previous resummation
calculation in Ref.~\cite{Brambilla:2018tyu} (BCK18) 
and the fixed-order calculation at two-loop accuracy in 
Ref.~\cite{Feng:2017hlu} (FJS17). In the case of $\eta_c$, 
we also show the result from a lattice QCD study in
Ref.~\cite{Colquhoun:2023zbc} (HPQCD23), the resummation calculation in 
Ref.~\cite{Bodwin:2001pt} (BC01), and the PDG value from
Ref.~\cite{Workman:2022ynf}. 
Note that the FJS17 results do not include any uncertainties from
uncalculated corrections of higher orders in $v$, leading to a severe tension
with the PDG result in the $\eta_c$ case. 
Generally, the improved results in this work lead to smaller values of the
branching fractions, which makes the results for $\eta_c$ 
agree better with the PDG value. 

Similarly to the $\chi_{QJ}$ case, we also list the results for the color-octet 
matrix elements $\langle \eta_Q | {\cal O}_8 (^3S_1) | \eta_Q \rangle$ 
and $\langle \eta_Q | {\cal O}_8 (^1P_1) | \eta_Q \rangle$ computed from the
expressions in Eqs.~(\ref{eq:etaQoctet}) and our results for ${\cal B}_1$ and
${\cal E}_1$. We compute the color-singlet matrix element 
from the expression $\langle \eta_Q | {\cal O}_1 (^1S_0) | \eta_Q
\rangle = 2 N_c |\Psi_{\eta_Q}(0)|^2$, where $\Psi_{\eta_Q}(r)$ is the
wave function of the $\eta_Q$ state, and the first-principles calculation of the 
$S$-wave quarkonium wave functions in Ref.~\cite{Chung:2020zqc}. We obtain 
\begin{subequations}
\label{eq:etaQoctetMEs1}
\begin{align}
\langle {\cal O}_8 (^3S_1) \rangle_{\eta_c(1S)} 
&= 
(-2.74^{+0.32}_{-0.40}) \times 10^{-2}~{\rm GeV}^3,
\\
\langle {\cal O}_8 (^3S_1) \rangle_{\eta_c(2S)} &= 
(-2.17^{+0.25}_{-0.32}) \times 10^{-2}~{\rm GeV}^3,
\\
\langle {\cal O}_8 (^3S_1) \rangle_{\eta_b(1S)} &= 
(-1.73^{+0.20}_{-0.25}) \times 10^{-2}~{\rm GeV}^3,
\\
\langle  {\cal O}_8 (^3S_1) \rangle_{\eta_b(2S)} &= 
(-1.01^{+0.12}_{-0.15}) \times 10^{-2}~{\rm GeV}^3,
\\
\langle  {\cal O}_8 (^3S_1) \rangle_{\eta_b(3S)} &=
(-0.84^{+0.10}_{-0.12}) \times 10^{-2}~{\rm GeV}^3,
\end{align}
\end{subequations}
\begin{subequations}
\label{eq:etaQoctetMEs2}
\begin{align}
|\langle {\cal O}_8 (^1P_1) \rangle_{\eta_c(1S)} | &\lesssim
6.5 \times 10^{-3}~{\rm GeV}^5, 
\\
|\langle {\cal O}_8 (^1P_1) \rangle_{\eta_c(2S)} | &\lesssim
5.1 \times 10^{-3}~{\rm GeV}^5, 
\\
|\langle {\cal O}_8 (^1P_1) \rangle_{\eta_b(1S)} | &\lesssim
5.3 \times 10^{-2}~{\rm GeV}^5, 
\\
|\langle {\cal O}_8 (^1P_1) \rangle_{\eta_b(2S)} | &\lesssim
3.1 \times 10^{-2}~{\rm GeV}^5, 
\\
|\langle {\cal O}_8 (^1P_1) \rangle_{\eta_b(3S)} | &\lesssim
2.6 \times 10^{-2}~{\rm GeV}^5. 
\end{align}
\end{subequations}
Here we again use the shorthand 
$\langle {\cal O}_8 (^3S_1) \rangle_{\eta_Q}
\equiv \langle \eta_c | {\cal O}_8 (^3S_1) | \eta_Q\rangle$
and $\langle {\cal O}_8 (^1P_1) \rangle_{\eta_Q}
\equiv \langle \eta_c | {\cal O}_8 (^1P_1) | \eta_Q\rangle$. 
In obtaining these results, we included the order-$\alpha_s$ corrections to the
wave function arising from loop corrections to the static potential, but
neglected corrections of order $\alpha_s^2$ because short-distance coefficients
for the color-octet channel are generally known up to next-to-leading order
in $\alpha_s$. In the case of 
$\langle \eta_Q | {\cal O}_8 (^1P_1) | \eta_Q \rangle$, 
we only show the upper estimates of the absolute values because 
our result for ${\cal E}_1$ is not precise enough to determine its sign; 
nevertheless the upper estimates in Eq.~(\ref{eq:etaQoctetMEs2}) are 
smaller than the power-counting estimates in Ref.~\cite{Bodwin:2002cfe}.

Next, we consider the decays of spin-triplet $S$-wave quarkonium $V$, 
which includes $J/\psi$ and $\Upsilon$. 
The following relations hold for the color-octet matrix elements for the 
$V$ state and the dimension-two moments~\cite{Brambilla:2002nu}:
\begin{subequations}
\label{eq:Jpsioctet}
\begin{align}
\label{eq:1S0relation}
\frac{\langle V | {\cal O}_8 (^1S_0) | V \rangle}
{\langle V | {\cal O}_1 (^3S_1) | V \rangle}
&= - \frac{c_F^2 {\cal B}_1}{6 m_Q^2 N_c},
\\
\label{eq:3PJrelation}
\frac{\langle V | {\cal O}_8 (^3P_J) | V \rangle/m_Q^2}
{\langle V | {\cal O}_1 (^3S_1) | V \rangle}
&= -(2 J+1) \frac{{\cal E}_1}{18 m_Q^2 N_c},
\end{align}
\end{subequations}
where $J=0$, 1, and 2. 
We again take the definitions of the NRQCD operators 
${\cal O}_8 (^1S_0)$, ${\cal O}_1 (^3S_1)$, and ${\cal O}_8 (^3P_J)$ 
in Refs.~\cite{Bodwin:1994jh, Bodwin:2002cfe, Brambilla:2002nu}. 
Note that these relations can also be obtained from Eqs.~(\ref{eq:etaQoctet})
by using heavy-quark spin symmetry (see Ref.~\cite{Bodwin:2002cfe}). 
Similarly to the $\eta_Q$ case, we do not consider the matrix element 
$\langle V | {\cal O}_8 (^3S_1) | V \rangle$, because it is proportional to a
moment of the four-point field-strength correlator, which is outside of the
scope of this work. 
The values of these color-octet matrix elements can be computed in the same way
as the $\eta_Q$ matrix elements, by using 
$\langle V | {\cal O}_1 (^3S_1) | V \rangle = 2 N_c | \Psi_V (0)|^2$ and the
first-principles calculations of the quarkonium wave functions $\Psi_V (r)$ in
Ref.~\cite{Chung:2020zqc}. We obtain 
\begin{subequations}
\label{eq:VoctetMEs1}
\begin{align}
\langle {\cal O}_8 (^1S_0) \rangle_{J/\psi} &=
(-9.1^{+1.1}_{-1.3}) \times 10^{-3}~{\rm GeV}^3,
\\
\langle {\cal O}_8 (^1S_0) \rangle_{\psi(2S)} &=
(-7.2^{+0.8}_{-1.1}) \times 10^{-3}~{\rm GeV}^3,
\\
\langle {\cal O}_8 (^1S_0) \rangle_{\Upsilon(1S)} &=
(-5.8^{+0.7}_{-0.8}) \times 10^{-3}~{\rm GeV}^3,
\\
\langle {\cal O}_8 (^1S_0) \rangle_{\Upsilon(2S)} &=
(-3.4^{+0.4}_{-0.5}) \times 10^{-3}~{\rm GeV}^3,
\\
\langle {\cal O}_8 (^1S_0) \rangle_{\Upsilon(3S)} &=
(-2.8^{+0.3}_{-0.4}) \times 10^{-3}~{\rm GeV}^3,
\end{align}
\end{subequations}
\begin{subequations}
\label{eq:VoctetMEs2}
\begin{align}
\frac{|\langle {\cal O}_8 (^3P_J) \rangle_{J/\psi}|}{2 J+1}
&\lesssim
7.2 \times 10^{-4}~{\rm GeV}^5,
\\
\frac{|\langle {\cal O}_8 (^3P_J) \rangle_{\psi(2S)}|}{2 J+1} 
&\lesssim
5.7 \times 10^{-4}~{\rm GeV}^5,
\\
\frac{|\langle {\cal O}_8 (^3P_J) \rangle_{\Upsilon(1S)}|}{2 J+1} 
&\lesssim
5.9 \times 10^{-3}~{\rm GeV}^5,
\\
\frac{|\langle {\cal O}_8 (^3P_J) \rangle_{\Upsilon(2S)}|}{2 J+1} 
&\lesssim
3.4 \times 10^{-3}~{\rm GeV}^5,
\\
\frac{|\langle {\cal O}_8 (^3P_J) \rangle_{\Upsilon(3S)}|}{2 J+1} 
&\lesssim
2.8 \times 10^{-3}~{\rm GeV}^5.
\end{align}
\end{subequations}
Again we use the shorthand 
$\langle {\cal O}_8 (^1S_0) \rangle_{V} \equiv
\langle V | {\cal O}_8 (^1S_0) | V \rangle$ and 
$\langle {\cal O}_8 (^3P_J) \rangle_V \equiv
\langle V | {\cal O}_8 (^3P_J) | V \rangle$. 
These results are same as what could be obtained from the $\eta_Q$ matrix
elements in Eqs.~(\ref{eq:etaQoctetMEs1}) and (\ref{eq:etaQoctetMEs2}) 
and heavy-quark spin symmetry. Similarly to the $\eta_Q$ case, we only show the
upper estimates of the absolute values of 
$\langle V | {\cal O}_8 (^3P_J) | V \rangle$. 

The results for the matrix elements we obtain 
can be used to estimate sizes of the color-octet
contributions to the decay rates $\Gamma(V \to {\rm LH})$. 
As have been known from Refs.~\cite{Petrelli:1997ge, Bodwin:2002cfe}, 
many of the short-distance coefficients associated with the color-octet matrix
elements are enhanced by a factor of $\pi/\alpha_s$, because a heavy-quark
antiquark pair in color-octet states can decay into two gluons or a light-quark
pair at tree level, while a color-singlet $S$-wave spin-triplet state can only
decay into three or more gluons at tree level. 
For example, the size of the color-octet contribution from the matrix element 
$\langle V | {\cal O}_8 (^1S_0) | V \rangle$ to the decay rate 
$\Gamma(V \to {\rm LH})$ relative to the one at leading order in $v$ is given
by $11.64 \times (\pi/\alpha_s)$ times
Eq.~(\ref{eq:1S0relation})~\cite{Bodwin:2002cfe}. 
Similarly, the contributions from the matrix elements 
$\langle V | {\cal O}_8 (^3P_0) | V \rangle$ and 
$\langle V | {\cal O}_8 (^3P_2) | V \rangle$ relative to the leading-order 
decay rate are given by Eq.~(\ref{eq:3PJrelation}) times 
$34.93 \times (\pi/\alpha_s)$ and $9.3 \times (\pi/\alpha_s)$,
respectively\footnote{ 
Exceptionally, the short-distance coefficient associated with the matrix
element $\langle V | {\cal O}_8 (^3P_1) | V \rangle$ is not enhanced by 
inverse powers of $\alpha_s$, because a vector state cannot decay into two
gluons at tree level.}. 
Due to the large coefficients, the color-octet contributions are significant
for $\Gamma(V \to {\rm LH})$, and can even exceed the color-singlet
contribution at leading order in $v$ in the charmonium case. 
For example, for $J/\psi$ the correction from ${\cal B}_1$ is about $-3.1$
times the leading-order decay rate, and the correction from ${\cal E}_1$ 
ranges from $-1.0$ to $+0.6$ times the leading-order decay rate. 
The size of the contribution from the matrix element 
$\langle J/\psi | {\cal O}_8 (^3S_1) | J/\psi \rangle$ 
that involves the four-point correlation function 
would also be order one if we use the estimate for the matrix element
given in Ref.~\cite{Bodwin:2002cfe}. 
Even in the case of $\Upsilon$, the color-octet contributions from 
$\langle \Upsilon | {\cal O}_8 (^1S_0) | \Upsilon \rangle$
and $\langle \Upsilon | {\cal O}_8 (^3S_1) | \Upsilon \rangle$
can be as large as 50\% of the leading-order decay rate, 
although the correction from 
$\langle \Upsilon | {\cal O}_8 (^3P_J) | \Upsilon \rangle$
is expected to be at most $\pm0.15$ times the leading-order decay rate. 
Since the contributions from ${\cal B}_1$ to the decay rate is negative, 
it is possible that large cancellations may occur between color-octet
contributions. Nonetheless, even in the case of $\Upsilon$ decays, knowledge of
the matrix element $\langle V | {\cal O}_8 (^3S_1) | V \rangle$ 
would be necessary to compute the decay rate to some precision. 

Aside from the exceptional case of the decays of spin-triplet $S$-wave
quarkonium into light hadrons, 
the moment ${\cal E}_1$ may be neglected in most cases due to its tiny size. 
For example, ${\cal E}_1$ appears in order-$v^2$ corrections to the
electromagnetic production and decays of $\chi_{QJ}$ (see
Refs.~\cite{Brambilla:2002nu, Brambilla:2020xod}). 
The correction from ${\cal E}_1$ to the two-photon decay amplitude 
of $\chi_{QJ}$ is given by $\frac{4}{9} {\cal E}_1 /m_Q^2$ for $J=0$ and 
$\frac{2}{3} {\cal E}_1 /m_Q^2$ for $J=2$
relative to the leading-order result~\cite{Brambilla:2020xod}.
These are at most of the order of 5\% compared to the tree-level
leading-order result for the charmonium case, 
and less than 1\% relative to the leading-order result for the bottomonium case, 
so they may be neglected in the phenomenology of $P$-wave quarkonium decays
at the current level of accuracy. 

Finally, we discuss phenomenological applications of $i {\cal E}_2$. 
This quantity appears in the correction to the relation between the
color-singlet matrix element and the wave function at the origin for $P$-wave
states [Eq.~(\ref{eq:chiQsinglet})], which was computed in
Ref.~\cite{Brambilla:2020xod}. This correction was later found to be
canceled by the correction to the $P$-wave wave function at the origin coming
from the velocity-dependent potential at zero separation (see
Ref.~\cite{Chung:2021efj}). Hence, $i {\cal E}_2$ does not have a
phenomenological significance in $P$-wave quarkonium decays or production. 
However, $i {\cal E}_2$ does appear in the correction to the wave function at
the origin for $S$-wave quarkonia that arises from the velocity-dependent
potential at zero separation~\cite{Chung:2020zqc, Chung:2021efj}, 
which unlike the $P$-wave case is not canceled by corrections to the relations 
$\langle V | {\cal O}_1 (^3S_1) | V \rangle = 2 N_c | \Psi_V (0)|^2$
and $\langle \eta_Q | {\cal O}_1 (^1S_0) | \eta_Q
\rangle = 2 N_c |\Psi_{\eta_Q}(0)|^2$.
The correction to the $S$-wave wave function at the origin $\Psi (0)$ 
relative to the leading-order result is given by 
$- i {\cal E}_2/(3 m_Q)$, which can be obtained from the calculation in 
Ref.~\cite{Chung:2020zqc} and the identification of the velocity-dependent
potential at zero separation in terms of $i {\cal E}_2$ in 
Ref.~\cite{Chung:2021efj}. 
In Ref.~\cite{Chung:2020zqc} this correction was considered only in the
uncertainties, which were estimated by assuming that $| 2 i {\cal E}_2/3 |
\lesssim 0.5$~GeV. This estimate is more than twice as large compared to the
central value of our result for $i {\cal E}_2$ in Eq.~(\ref{eq:E2_lat}), 
and hence, our result for $i {\cal E}_2$ can be used to significantly reduce
this uncertainty. 
By including the correction from the velocity-dependent potential at zero 
separation, the $S$-wave charmonium wave functions
at the origin are enhanced by $7\pm 4$\%, and the $S$-wave bottomonium
wave functions at the origin are enhanced by $2 \pm 1$\%. 
In Ref.~\cite{Chung:2020zqc}, the uncertainty from the unknown $i {\cal E}_2$
was estimated to be about 19\% and 5\% for $S$-wave charmonium and bottomonium
wave functions at the origin, respectively; by using our result for $i {\cal
E}_2$, these uncertainties would be greatly reduced to only about $4\%$ and 
$1\%$, respectively. 
For example, using our result for $i {\cal E}_2$ would improve 
the first-principles calculation of the $J/\psi$ leptonic
decay rate in Ref.~\cite{Chung:2020zqc} from 
$4.5^{+2.5}_{-1.9}$ to $5.1^{+1.6}_{-1.4}$~keV, which not only reduces
uncertainties but also brings the central value closer to the PDG value 
$5.53 \pm 0.10$~keV~\cite{Workman:2022ynf}. 
In the case of the two-photon decay rate of $\eta_b(1S)$, we obtain only a 
small improvement in uncertainty, from $0.433^{+0.165}_{-0.065}$ to 
$0.450^{+0.149}_{-0.047}$~keV. 
When combined with our results for $R_{\eta_b}$ in Eq.~(\ref{eq:Retab}), we
obtain the total decay rate of $\eta_b (1S)$ given by
$11.3^{+3.8}_{-1.2}$~MeV, which is slightly improved compared to
Eq.~(\ref{eq:etabtotalrates}).

\section{Summary and Discussion}
\label{sec:conclusion}

In this work we computed color-octet nonrelativistic QCD matrix elements that
appear in heavy quarkonium decay rates based on the refined Gribov-Zwanziger
theory~\cite{Dudal:2007cw, Dudal:2008sp, Dudal:2008rm, Dudal:2008xd,
Dudal:2010tf, Dudal:2011gd}. The color-octet matrix elements correspond to the
probabilities for a heavy quark and antiquark pair in a color-octet state to
evolve into a color-singlet state, and can be related to the moments of
correlation functions of QCD field-strength tensors by using the potential
nonrelativistic QCD formalism~\cite{Pineda:1997bj, Brambilla:1999xf,
Brambilla:2001xy, Brambilla:2002nu, Brambilla:2004jw}. 
We found that tree-level calculations in the refined Gribov-Zwanziger theory, 
which can reproduce the nonperturbative gluon propagator very well, can also 
give a satisfactory description of the two-point correlation functions of the
QCD field-strength tensor that agrees with a lattice QCD study in
Ref.~\cite{Bali:1997aj}. 
This allowed us to compute moments of the correlation functions that are
infrared finite, while also reproducing the correct leading ultraviolet 
divergences that are expected from the nonrelativistic QCD factorization
formalism~\cite{Bodwin:1994jh} and can be properly renormalized. 
In particular, the dimensionless moment ${\cal E}_3$ defined in
Eq.~(\ref{eq:E3def}) contains a logarithmic ultraviolet divergence, 
which we renormalize in the $\overline{\rm MS}$ scheme consistently with
calculations in the nonrelativistic QCD factorization formalism.
Similarly, in the case of the dimension-two moments ${\cal E}_1$ and ${\cal B}_1$
defined in Eq.~(\ref{eq:E1B1defs}), which involve quadratic power divergences, 
correct results in dimensional regularization are obtained by computing
them in arbitrary spacetime dimensions and later setting the number of
dimensions to 4. Such a calculation in dimensional regularization would be
difficult to carry out in a numerical study, such as a calculation based on
lattice QCD data, especially for the dimensionless moment ${\cal E}_3$ where
the conversion to dimensional regularization would involve not only 
ultraviolet divergences but also infrared divergences. 

The numerical result for ${\cal E}_3$ that we obtain in the refined
Gribov-Zwanziger theory is given in Eq.~(\ref{eq:E3result_1GeV_final}). 
This result is much more
precise than the results from phenomenological analyses in
Refs.~\cite{Brambilla:2001xy, Brambilla:2020xod}. 
We have used this result to compute the color-octet matrix element for
$P$-wave quarkonium decays and to compute decay rates of $\chi_{cJ}$ and
$\chi_{bJ}$ into light hadrons. 
The uncertainties that arise from color-octet matrix element have significantly
improved compared to previous phenomenological studies, 
although the results are generally compatible with previous results and
experimental data. 

In the case of the dimension-two moment ${\cal B}_1$, we found that a
straightforward calculation in the refined Gribov-Zwanziger theory leads to 
not only a power ultraviolet divergence, as expected in perturbative QCD, but
also a logarithmic ultraviolet divergence that is proportional to a
nonperturbative parameter associated with the dimension-two condensate of the
gauge field. While appearance of subdivergences is not completely unexpected 
given that the refined Gribov-Zwanziger theory contains dimensionful
parameters,
due to the nonperturbative origin of the subdivergences, they could not be
subtracted within the nonrelativistic QCD factorization formalism. 
Fortunately, we found that at tree level, this result is exactly proportional
to the dimension-two condensate of the gauge field which also contains a
logarithmic ultraviolet divergence as was found in
Refs.~\cite{Verschelde:2001ia, Dudal:2011gd}. By using the finite result for
the condensate obtained in the effective action formalism in
Ref.~\cite{Verschelde:2001ia}, we obtain a finite result for ${\cal B}_1$ in
dimensional regularization, which is given in Eq.~(\ref{eq:B1RGZ_numberfinal}). 
This result lead to the calculation of the color-octet matrix elements 
that appear in $S$-wave quarkonium decays associated with the spin-flip
interaction. 
We have used this result to improve a previous calculation of the two-photon
branching fraction of $\eta_c$ and $\eta_b$ in Ref.~\cite{Brambilla:2018tyu}, 
where the color-octet matrix elements have been left unknown. 
By including the color-octet matrix elements computed from the calculation of 
${\cal B}_1$ in the refined Gribov-Zwanziger theory, the central values of the
branching fractions have increased, and the uncertainties associated with the
color-octet matrix elements have reduced. 

Finally, we found that the dimension-two moment ${\cal E}_1$ vanishes at tree
level in the refined Gribov-Zwanziger theory. 
We also found that in general, 
the contribution from the invariant function ${\cal D}(x^2)$ vanishes in 
${\cal E}_1$, and only ${\cal D}_1(x^2)$ contributes to ${\cal E}_1$, which
in perturbation theory appears from next-to-leading order in the strong
coupling. 
We have estimated the size of the contribution from ${\cal D}_1(x^2)$
to ${\cal E}_1$ by using lattice QCD measurements of the invariant functions in
Ref.~\cite{Bali:1997aj} to obtain the result in Eq.~(\ref{eq:E1_RGZ}). 
The result suggests that ${\cal E}_1$ is at most about an order of magnitude
smaller than ${\cal B}_1$, and so, in most phenomenological applications, 
contributions from ${\cal E}_1$ may be neglected, unless it is multiplied by an
unusually large short-distance coefficient. 

The results for the two-point field-strength correlation function in this work
are based on tree-level calculations of the gluon propagator in the refined
Gribov-Zwanziger theory.
Similarly to the case of the gluon propagator, tree-level results in this
theory seem to be able to reproduce bulk of the long-distance nonperturbative
behavior of the correlation function, as we find agreements with the lattice
QCD analysis in Ref.~\cite{Bali:1997aj}. 
We also found that the Gribov-Zwanziger theory alone~\cite{Zwanziger:1989mf}, 
without the refinements from the dimension-two condensates, 
cannot reproduce the long-distance behavior found in lattice studies. 
This suggests that inclusion of the dimension-two condensates is necessary in 
describing the nonperturbative nature of the field-strength correlation
functions, similarly to the case of the gluon propagator. 
Although the results in this work are based on calculations at tree level, 
in obtaining the numerical results we included effects of contributions at
higher orders in the strong coupling estimated by comparing with lattice QCD
results. 
It would still be interesting to compute the correlation functions at
next-to-leading order accuracy, which may also help reduce the uncertainties in
the numerical results. 
We also note that in this work we have not computed the color-octet matrix
element that arises from the four-point field-strength correlation function. 
In principle, we could compute moments of the four-point correlation
function at tree level similarly to our calculation of the two-point 
correlation function.  However, we found that such a calculation leads to
logarithmic ultraviolet divergences, which, unlike the case of ${\cal B}_1$, 
involve not only $m^2$ but also $M^2$ and $\lambda$, which makes it 
impossible to reexpress the divergence in terms of the dimension-two
condensate of the gluon field. Extending the analysis to the case of the 
four-point correlation function and also to next-to-leading order accuracy 
would be important in the phenomenology of $J/\psi$ and $\Upsilon$ decays, 
as the short-distance coefficients associated with color-octet matrix elements
are very large for these processes. 

\acknowledgments 

This research was supported by Basic Science Research Program through the
National Research Foundation of Korea (NRF) funded by the Ministry of
Education (Grant No. RS-2023-00248313),
the National Research Foundation of Korea
(NRF) Grant funded by the Korea government (MSIT) under Contract No.
NRF-2020R1A2C3009918, and by a Korea University grant.

\bibliography{qoctetGZ.bib}

\end{document}